\documentclass[11pt,a4paper]{article}
\usepackage{jheppub}
\usepackage[utf8]{inputenc}
\usepackage[english]{babel}
\makeatletter
\usepackage{graphicx}
\usepackage{epstopdf}
\usepackage{mathtools}
\usepackage[strict]{changepage}
\usepackage[utf8]{inputenc}
\usepackage{xcolor}
\usepackage{subcaption}
\usepackage{hyperref}
\usepackage{mathrsfs}
\usepackage{amsthm}
\usepackage{tikz-feynman}
\usepackage[makeroom]{cancel}

\newtheorem{definition}{Definition}

\begin{document}

\title{Interplay between reflection positivity and crossing symmetry in the bootstrap approach to CFT}

\author[a]{Leandro Lanosa}
\author[b]{Mauricio Leston}
\author[c]{Mario Passaglia}
\affiliation[a]{Departamento de Matem\'atica - FCEyN - Universidad de
Buenos Aires and IMAS - CONICET, Pabell\'on I, Ciudad
Universitaria, C1428EHA Buenos Aires, Argentina.}
\affiliation[b]{\it Instituto de Astronom\'ia y F\'isica del Espacio (IAFE - CONICET), Universidad de Buenos Aires,
1428 Buenos Aires, Argentina.}
\affiliation[c]{Departamento de Física, Universidad de Buenos Aires,
	1428 Buenos Aires, Argentina.}
\emailAdd{lanosalf@gmail.com, mauricio@iafe.uba.ar,mario.passaglia@gmail.com}

\date{\today}

\abstract{Crossing symmetry (CS) is the main tool in the bootstrap program applied to CFT. This consists in an equality which imposes restrictions on the CFT data of a model, i.e., the OPE coefficients and the conformal dimensions. Reflection positivity (RP) has also played a role in this program, since this condition is what leads to
the unitary bound and reality of the OPE coefficients. In this paper, we show that RP can still reveal more information, explaining how RP itself can capture an important part of the restrictions imposed by the full CS equality. In order to do that, we use a connection used by us in a previous work between RP and positive definiteness of a function of a single variable. This allows us to write constraints on the OPE coefficients in a concise way . These constraints are encoded in the conditions that certain functions of the cross-ratio will be positive defined and in particular \textit{completely monotonic}. We will consider how the bounding of scalar conformal dimensions and OPE coefficients arise in this RP based approach. We will illustrate the conceptual and practical value of this view trough examples of general CFT models in $d$-dimensions.}

\keywords{CFT, Reflection positivity, Bootstrap}

\maketitle

\section{Introduction}\label{intro}

The aim of the bootstrap program in quantum field theories with conformal invariance (CFT) (see \cite{Poland:2018epd} for a review) is to constraint as much as possible the CFT data: the set of conformal weights and operator product expansion (OPE) coefficients, i.e. the structure constants. In this program, crossing symmetry is the main ingredient since it provides some equalities involving the OPE coefficients, highly constraining them. Reflection positivity (RP), established as a rigorous theorem in \cite{Osterwalder:1973dx} and sometimes referred to as {\it unitarity}, has also played a role in this program. The most celebrated example of constraints imposed by RP is the so-called unitary bound for conformal dimensions of different operators in any space-time dimension. Another consequence of RP is the reality of OPE coefficients corresponding to real operators. 

In contrast with crossing symmetry, RP consists of a set of inequalities (infinitely many of them) involving the euclidean continuation of vacuum expectation values (Schwinger functions) smeared with test functions of several arguments. This set of inequalities tells that the Schwinger functions come from vacuum expectation values of operators acting in a Hilbert space which is naturally endowed with a positive definite bi-lineal form. The term {\it reflection} refers to a reflection operation $\Theta$ in the euclidean hyperplane. Roughly speaking, $\Theta$ changes the sign of the coordinate $x_0$ corresponding to the temporal variable in the Lorentzian formulation of the CFT. RP inequalities arise by taking the square norm of certain states in the Lorentzian CFT.

It is natural to ask whether, besides the known restrictions (unitary bounds and the reality of OPE coefficients), there are more constraints coming from RP. In other words, if we restrict ourselves to the set of CFT constrained by these two restrictions, is RP consistent with any CFT data there? A definitive answer will require a proof that, for any conformal dimensions and OPE coefficients fulfilling the mentioned restrictions, RP inequalities are automatically fulfilled. Regarding this, there was a lot of work, specially during the seventies, for instance the early works of G. Mack \cite{Mack:1974sa}. This issue was considered recently in \cite{RychkovRP}. There, it was proposed certain minimalist set of axioms, which include the RP of 2-point function for primary and descendants fields (which is equivalent to unitary bound) and the condition that the OPE coefficients must be real. From this set of axioms, it was given a proof of RP at level of 4-points functions\footnote{As it is mentioned there, the complete proof of RP for higher order $n$-points functions will require to overcome technical difficulties in the OPE expansion or to make stronger some of the proposed axioms.}.

Therefore, it is clear that RP, at least at the level of 4-points functions, will not lead to any further constraints on the CFT data. In fact, one can check this in the particular example of 4-points functions involving equal spinless fields $\phi$ of dimension $\Delta_0$. Here, we consider this case in order to keep the discussion simple. The 4-point function ${<0\mid\phi(x_1)\phi(x_2)\phi(x_3)\phi(x_4)\mid{0}>}$ is, up to a factor, equal to a function $G(x)$ of the cross-ratio $x$, which is real for the case of aligned points. This function admits a decomposition in terms of conformal blocks $F$ as follows:
$$G(x)= x^{-2\Delta_0}\sum_{p}{(C_{\Delta_0\Delta_0}^p)}^2 F^{(p)}(x)$$

where $p$ stands for the pair $(\Delta, \ell)$ (conformal dimension and spin) of all the exchange fields and the $F$'s are the conformal blocks, completely fixed by general properties of the CFT. It turns out that the conformal blocks $F$ themselves fulfill the requirements of RP and since they are combined with positive coefficients (the $C$'s are real) the 4-points function is automatically RP. Therefore, RP does not impose restriction on the $C's$, as expected from \cite{RychkovRP}.
	
However, there is a way in which RP can be used to see some of the constraints on the OPE coefficients implied by CS. In our simple example of equal spinless fields, CS states that $G$ is symmetric around $x=\frac{1}{2}$:
\begin{equation}
G(x) = G(1-x)\label{CS},
\end{equation}

In a more general case of non-equal fields, the function on the lhs is different from the one on the rhs. But let us focus now on this simple case.
The standard approach consists in finding ``forbidden regions'' for the conformal dimensions of the fields appearing in the equation, i.e., it consists in finding the regions where the equation has not solutions for real OPE coefficients. The main idea is to demand that all odd derivatives of $G$ vanish at $x=\frac{1}{2}$ (the procedure is slightly different). These derivatives are obtained by summing all the contributions of each conformal block. This is the so-called ``sum rule", introduced in \cite{RychkovBound} and subsequently improved (see for instance \cite{Navegator} for the progress in the numerical techniques)

The idea of the present paper comes from the observation that the lhs of (\ref{CS}), as a function of the cross-ratio $x$, obeys some inequalities inherited from RP and therefore the rhs, as function of $x$, should also obey those inequalities. Let us notice that, even in this simplified case where the same $G$ appears at both sides, on the rhs it appears the composition of $G$ with $1-x$. This change of $x$ by $1-x$ implies that, on the rhs, these inequalities are not automatically fulfilled, and at least the OPE coefficients are restricted. In order to understand this, let us briefly describe the connection between RP and some inequalities for a function of a single variable used by us in \cite{Blanco:2019gmt}.

In that paper we showed that Rindler positivity (a positivity conditions similar to RP) leads to the more simple property of positive definiteness (pd) of a single variable function. Pd functions have nice properties which can be translated into a positivity of certain coefficients. In particular, when a pd function $f$ on $(a,+\infty)$ is bounded, it can be proved that $f$ should be a {\it a complete monotonic (CM) function}, i.e., a function fulfilling $(-1)^k f^{(k)}(x)\geq{0}$. By choosing sharp test functions located in certain one-parameters family of points, one can see that RP implies that, up to a factor, $G$ has to be a CM function of both $\rho\equiv{-\log(x)}$ and $\eta \equiv -1 + 1/\sqrt{x}$.

Let us explain in a more precise way the strategy that we will use in this paper for the $\rho$ case first. As we have said before, according to \cite{RychkovRP}, RP is automatically fulfilled at the level of the 4-points function. It means that the lhs of eq (\ref{CS}) should be automatically a CM function of $\rho=-\log(x)$ , regardless of what values the real OPE coefficients take. In fact, the conformal blocks $F$ are CM functions of $\rho$ for any exchange values $(\Delta, \ell)$, as it can be checked in several examples that we will see later. Since they are combined with positive coefficients (the $C$'s are real) the lhs of eq (\ref{CS}) fulfills automatically RP, not imposing restrictions on the $C's$, as expected from \cite{RychkovRP}. However, on the rhs, due to the substitution of $x$ by $1-x$, the conformal blocks $F$ lose generically its CM property. We will see that, for some region in $(\Delta, \ell)$ space of exchange fields, the functions accompanying the square of the structure constants in the rhs are not CM functions. This means that RP is not automatically guaranteed on the rhs for arbitrary OPE coefficients. This is the core of our strategy: the OPE coefficients should be constrained in order to make the rhs a CM function of $\rho$ (and similarly for $\eta$).

These constrains on the OPE coefficients, deduced using the rhs CM property, will not be stronger than those coming from the CS equality itself. In other words, if we solve the CS equality, the rhs will be a CM function of $\rho$ (and of $\eta$), so the constrains arising from the mentioned combination of RP and CS are of course already included in the full constrains imposed by the CS eq. (\ref{CS}). However, this interplay between RP and CS allows us to have a rough idea of some of the constraints on the space of OPE coefficients which could be generally more difficult to see in the full program. As we will see, one can get pictures like the one of Figure \ref{Fig:newbound4d} of section \ref{linearRHS}. There, the upper zone contains those conformal dimensions contributing with the ``incorrect" sign (for being a CM function), enforcing the presence of at least one exchange field in the lower zone that contributes with the ``correct" sign in order to balance the positivity and ultimately leading to a CM rhs.

Since we will refer many times to the role of the lhs and rhs in the derivation of constraints, we want to emphasize that there are two distinct aspects of the difference between them. The first one has to do with the substitution of $x$ by $1-x$, whose relevance will be understood after we review the issue of pd functions of a single variable. The second difference appears in the more generic case in which we consider $4-$points functions corresponding to spinless fields with two non-equal conformal dimension. In that case, eq. (\ref{CS}) will be replaced by:
$$G^{21}_{21}(x)=G^{11}_{22}(1-x)$$

Then, there will be two different $G$'s at both sides in this case. As we will see, the lhs has a form directly related with a norm of a state, were RP applies, whereas the rhs doesn't.

We want to stress that this approach does not intend to compete with the standard ones based on the sum rule, which has already provided a lot of precise estimates on the conformal dimensions and OPE coefficients and that become more precise as far the computational methods were improved. Our work should rather be considered as a discussion on the compatibility of CS and RP, showing how (some consequences of) RP is manifest in the CS constraints and can be used as a shortcut for getting a global idea of the space of allowed conformal dimensions. In this paper we want to focus on simple cases in order to explain the main ideas, leaving more involved cases for a future work.

{\bf Organization of the paper}: In section \ref{RPrepaso}, we recall the general form of RP inequalities and the material about positive definite functions of single variables
with the one-parameter arrange of points used in \cite{Blanco:2019gmt}. In section \ref{linkblock} we will see the implications of RP in the case of CFT and check some positivity conditions in the conformal blocks. Then, in section \ref{CrossinPositivity}, after a warming up with the Ising model in $d=2$, we will present the main strategy of combining CS and RP, starting from simple examples. Finally, we add some comments about the potential use of this approach.

\section{RP inequalities}\label{RPrepaso}

In order to keep it simple, we will avoid the use of general test functions, which is the right way to deal with distributions. For simplicity also, we will consider the case of a 1+1 hermitian scalar fields.

We will use a particular case of RP inequalities as follows: let us consider sequences $X$ of ``length N'', made of $N$ time ordered points:

$$X=\{ (t_1,r_1), (t_2,r_2), ...(t_N,r_N)\},$$

with $0<t_{\alpha}<t_{\alpha+1}$; $r_i$ denote the spacial part. Let us consider $M$ sequences of this kind $X_i=\{x^{i}_{1},...x^{i}_{N}\}$, where we have added the supra index $i$ to each spacetime point. Using that the Hamiltonian of the QFT is bounded from below, one can define states of the form:

$$\psi_{i}=\prod_{\alpha=1}^Ne^{-H{(t^{(i)}_{\alpha}-t^{(i)}_{\alpha-1}})}\phi_\alpha (0,r^{(i)}_{\alpha})\Omega, $$ 

where $\phi_\alpha$ stands for differents scalar fields (in the case of a CFT each with conformal dimension $\Delta_\alpha$), $\Omega$ for the vacuum state and where we have set $t^i_{0}=0$.

The RP inequalities follows then by demanding that the square norm of $\psi=\sum_{i=1}^MC_{i}\psi_i$ ($C_i$ arbitrary complex numbers) will be positive:

\begin{equation*}
\lVert{\psi}\rVert^2=\sum_{i,j}C_iC_j^* (\psi_j,\psi_i)\geq{0} 
 \end{equation*}
 
There are an infinite number of inequalities of this type, since we are free to choose the sequences $X_i$ appearing in the sum and the coefficients $C_i$. Let us see how these inequalities look like in the particular case in which we only allow sequences $X_i$ of length 2:

\begin{equation*}
\begin{aligned}
\sum_{i,j}&C_iC_j^*(\psi_j,\psi_i)=\\
&\sum_{i,j}C_iC_j^*(\Omega, \phi_2(0,r^{j}_2)e^{-H(t_2^{j}-t_1^{j})}\phi_1(0,r^{j}_1)e^{-H(t^{i}_1+t^{j}_1)}\phi_1 (0,r^{i}_{1})e^{-H(t^{i}_2-t^{i}_1)}\phi_2 (0,r^{i}_2)\Omega)\geq{0}
\end{aligned}
\end{equation*}

We can see that $(\psi_j,\psi_i)$ is the Euclidean continuation of the Wightman function 
$$(\Omega, \phi_2 (-t^{j}_2,r^j_2)\phi_1 (-t^{j}_1,r^j_1)\phi_1 (t^{i}_1,r^i_1)\phi_2 (t^{i}_2,r^i_2)\Omega)$$

i.e, the Schwinger 4-point function $S_4$ evaluated at the points of the composed sequence $\Theta(X^j)X^i$, where $\Theta$ is defined by $\Theta{X}=\{(-t_N,r_N),(-t_{N-1},-r_{N-1},....(-t_{1},r_1)\}$. $\Theta$ makes a reflection in the time coordinate. With this notation, now in general for sequences $X_i$ of length $N$, RP states that for any set of complex numbers $C_i$ ($i=1,...,M$, $M$ arbitrary large), it holds:

\begin{equation}
\sum_{i=1}^M C^*_j C_i S_{2N}(\Theta(X_j)X_i)\geq{0}\label{RP}
\end{equation}

There are an infinite number of these inequalities. This kind of inequalities express the positive definiteness of the quadratic form $S_{2N}(\Theta(.)(.))$ acting on a sequence of time ordered points $X_i$.

To be more concrete, let us restrict to the case $N=2$, i.e, to RP for the Schwinger 4-point functions:

 \begin{equation}
 \sum_{i,j=1}^MC_j^*C_iS_4(\Theta(X_j)X_i)\label{cuatro}\geq{0}
 \end{equation}

 Although we have not used test functions, the choice of each sequence correspond formally to test functions sharp at the $t^i_{\alpha}$. 
 
 We want to convert the positive definiteness of this quadratic form into a positive definiteness of a function of a single variable, as we did in \cite{Blanco:2019gmt}. Let us make a brief summary of it below.

\subsection{Positive definiteness of functions of a single variable}\label{pd}

The characterization and the properties of positive definite functions of a single variable have been intensely studied in the early twentieth century, mainly by Schoenberg, Widder and Bernstein. Several known properties of the vacuum correlation functions in relativistic QFT are obtained by the applications of some of these results. In this section, we give some definitions and enunciate key theorems on positive definite functions that are relevant to our paper. For a complete study of these topics we refer the readers to \cite{Bernstein,Widder,Berg} (also see \cite{2016arXiv160804010J} for a brief account of the main theorems).

There are two notions of positive definiteness for a function of a single variable. The definition we use here is the following:

\begin{definition}\nonumber
	A real function $f:(a,b)\rightarrow{\mathbb R}$ is positive definite (pd) if, for any natural number $N$ and for any choice of points $\{x_i\}$ ($i=1,\,...,\,N$) with $x_i\in(a,b)$, the matrix $M$ of coefficients $M_{ij}\equiv{f}(\frac{x_i+x_j}{2})$ is positive definite \footnote{The other notion of positivity of a function arises when considering $M_{ij}=f(\vert{x}_i-x_j\vert)$ instead.}.
\end{definition}

Positive definiteness in this sense turns out to be a very restrictive condition. A surprising consequence of this property is the following: if $f:(a,b)\rightarrow{\mathbb R}$ is positive definite and continuous in $(a,b)$, then it is $C^{\infty}(a,b)$ (even more, it is real analytic there \cite{Widder}).

Moreover, the derivatives $f^{n}(x)$ of order $n$ satisfy an infinite set of inequalities valid at any $x\in(a,b)$: the $N\times{N}$ matrices ${H}^{(N,f)}$ of coefficients $\left({H^{(N,f)}}\right)_{m,n}=f^{(n+m)}$ ($n,m=0, ..., N-1$) are positive definite,
\begin{equation}
	\det H^{(N,f)}=\left|
	\begin{array}{ccccc}
		f & f^{(0+1)} & f^{(0+2)}&.. &f^{(0+N-1)}  \\
		f^{(1+0)} & f^{(1+1)} & .&. & .. \\
		.. & .. & . & .& \\
		f^{(N-1+0)} & . & .&. & f^{(N-1+N-1)} \\
	\end{array}
	\right|\geq{0}\label{det1}
\end{equation}

for all $N\in{\mathbb N}$. Conversely, an analytic function satisfying this infinite set of inequalities is pd. In fact, the inequalities (\ref{det1}) need only be satisfied at one point in $(a,b)$ and then they are automatically satisfied throughout the interval \cite{2016arXiv160804010J}.

An obvious consequence of the definition of positive definiteness is that a pd function is non-negative. A less obvious consequence is that the even derivatives of a pd function are also pd (this follows easily from the inequalities (\ref{det1})), and hence non-negative. Note also from the definition that a linear combination of pd functions with positive coefficients is also pd.

Simple examples of pd functions are $f(t)=e^{\lambda{t}}$ for $\lambda$ a real number. The positive definiteness can be checked easily both from the definition and from the inequalities (\ref{det1}). The definition of pd function requires that $\sum_{i,j=1..N}c_ic_jf(\frac{t_i+t_j}{2})\geq{0}$. In this case, $\sum_{i,j=1..N}c_ic_jf(\frac{t_i+t_j}{2})=(\sum_{i=1...N} c_ie^{\frac{\lambda{t}_i}{2}})^2\geq{0}$. Therefore, $f$ is pd. Checking the inequalities is trivial since all the determinants are just $0$. Linear combinations of exponential functions with positive coefficients will also be pd. In particular, a constant function $f(t)=c$, with $c\geq{0}$ is a pd function.

Pd functions are closely related to {\emph{absolutely monotonic}} (AM) and {\emph{completely monotonic}} (CM) functions, whose definitions are the following:
\begin{definition}
	A function $f$ is said to be absolutely monotonic (AM) if $f^{(n)}\ge 0$ for all $n=0,1,\dots$ and completely monotonic (CM) if $(-1)^nf^{(n)}\ge 0$ for all $n=0,1,\dots$.
\end{definition}
Note that the exponential $f(t)=e^{\lambda{t}}$ is AM for $\lambda>0$ and CM for $\lambda<0$. A pd function can always be written as the sum of an AM function and a CM function. This follows from a classical theorem (see \cite{Widder},Theorem VI.21), which states that a function $f$ on $(a,b)$ is pd if and only if it admits the following integral representation:
\begin{equation}\label{wieder}
	f(t)=\int _{-\infty}^\infty e^{-\lambda{t}}g(\lambda)d\lambda=\int _{-\infty}^0 e^{-\lambda{t}}g(\lambda)d\lambda + {\int _{0}}^{\infty} e^{-\lambda{t}}g(\lambda)d\lambda
\end{equation}

where $g$ is non-negative (strictly speaking, $g(\lambda)d\lambda$ has to be understood as a Borel measure). Note that the first term on the rhs side above is AM and the second term is CM.

\textit{Most important for this paper are the pd functions defined on $(0,+\infty)$ (or more generally on any interval of the form $(a,+\infty)$) which are bounded at infinity. From the decomposition (\ref{wieder}) it follows that such functions are necessarily CM}.

Roughly speaking, this is because the first term in (\ref{wieder}) diverges as $t\to\infty$, so this term must be absent in order for $f$ to be bounded at infinity (for a technical proof of this see \cite{Bernstein}). Conversely, it can be shown \cite{Widder} that any CM function on $(0,+\infty)$ admits the integral representation of the second term in (\ref{wieder}), and hence it is pd. In other words, {\it the space of pd functions on $(0,+\infty)$ which are bounded at infinity is equal to the space of CM functions on the same interval}.

This equivalence gives rise to additional inequalities to the ones given by equation (\ref{det1}), which come from the obvious fact that, if $f$ is CM, then $-f'$ is also CM. Using this and the above equivalence, we conclude that, for $f$ pd on $(0,+\infty)$ and bounded at infinity, $-f'$ is also pd.

The additional inequalities arise from substituting $f$ by $-f'$ in (\ref{det1}):

\begin{equation}
	\det H^{(N,-f')}=(-1)^N\left|
	\begin{array}{ccccc}
		f' & f'^{(0+1)} & f'^{(0+2)}&.. &f'^{(0+N-1)}  \\
		f'^{(1+0)} & f'^{(1+1)} & .&. & .. \\
		.. & .. & . & .& \\
		f'^{(N-1+0)} & . & .&. & f'^{(N-1+N-1)} \\
	\end{array}
	\right|\geq{0}\label{det2}
\end{equation}

Thus, pd functions on $(0,+\infty)$ which are bounded at infinity are characterized by two equivalent sets of conditions: (i) $(-1)^n f^{(n)}\geq{0}$ and (ii) equations (\ref{det1}) and (\ref{det2}). The first set of conditions appears to be much simpler than the second, but the second is more useful in some cases.

Having these tools in mind, let us come back to the issue of RP. What we will do next is to impose further restriction to the sequences $X_i$ mentioned in the previous subsection.

\subsection[RP applied to one-parameter aligned arrange of points in $d$ dimensions]{RP applied to one-parameter aligned arrange of points in $\boldsymbol{d}$ dimensions}\label{onepara}

Let us consider a sequence of aligned points in a euclidean $d$-dimensional space, in the direction of the coordinate ``$t$". Then, we can write: $X_i= \{t^{i}_1,t^{i}_2,...t^{i}_N\}$, with the time coordinate ordered as $0<t_{\alpha}<t_{\alpha+1}$. The main idea is to take a family of sequences $\{X_i\}$ such that, given a single real parameter $\lambda(i)$ that defines univocally the sequences, the quantity $S_{2N}(\Theta(X_j)X_i)$ will then be a function of a ``sum variable" $\lambda=\lambda(i)+\lambda(j)$. If that is possible for the chosen family, then we can apply the machinery for a positive definite function of a single variable $S_{2N}(\lambda)$ reviewed above. Restricting ourselves to length 2 sequences, there are two natural families (see Figure \ref{Fig:sequences}):

\begin{figure}\centering
\begin{tabular}{c|c}
\begin{subfigure}[b]{0.5\textwidth}
\includegraphics[width=\textwidth]{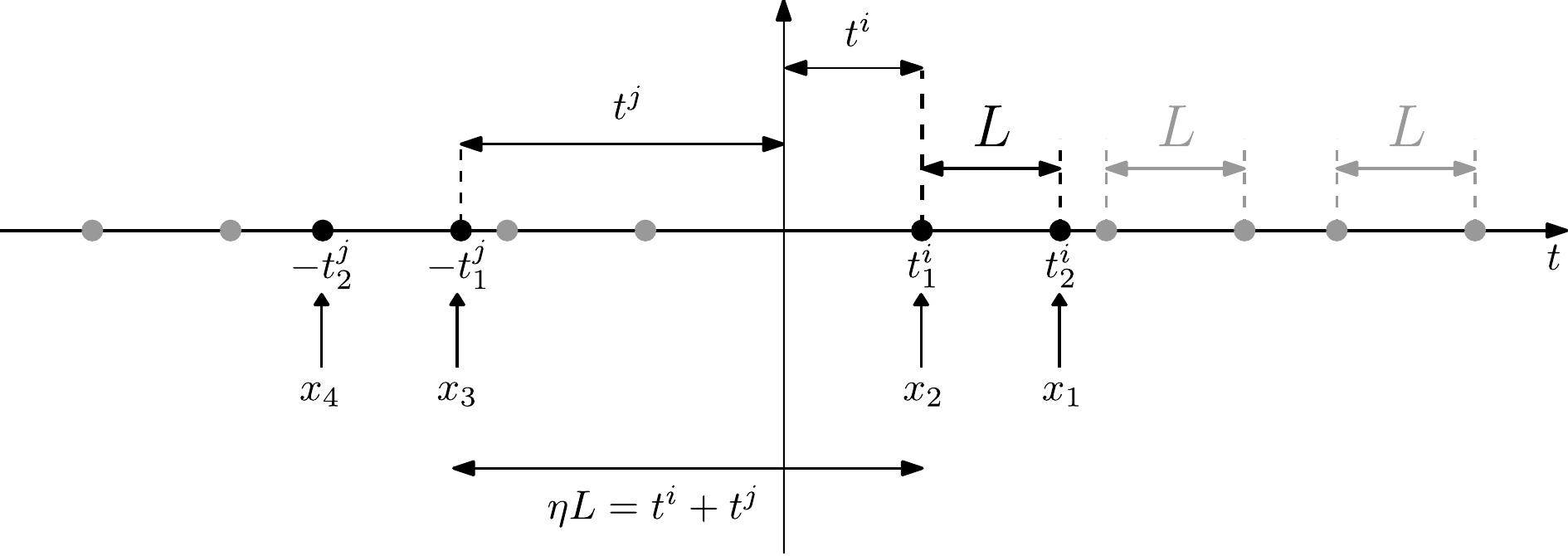}
\end{subfigure}
&
\begin{subfigure}[b]{0.45\textwidth}
\includegraphics[width=\textwidth]{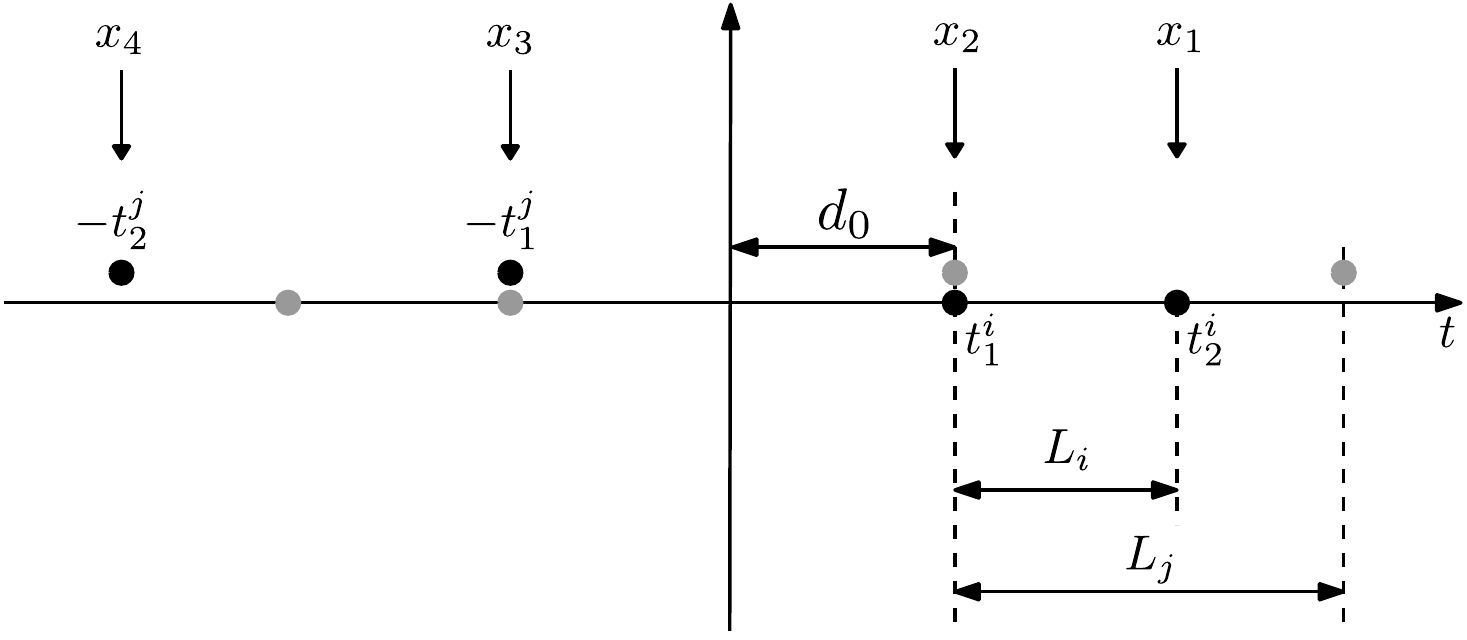}
\end{subfigure}
\end{tabular}
\caption{Left $\eta-$type, right $\rho-$type sequences}\label{Fig:sequences}
\end{figure}

\begin{enumerate}
\item {\bf First family} ($\eta-$type): $X_i=\{t^i,t^i+L\}$, $t^i>0,L>0$.
\item {\bf Second Family} ($\rho-$type): $X_i=\{d_0,d_0+L_i\}$, $d_0>0, L_i>0$
\end{enumerate}

The names refers to the parameters that will be introduced soon. In the first one, the two points of each sequence are separated by a fixed quantity $L$, and the single real variable that defines each sequence in this family is $t^i$.

In the second one, the separation is not fixed but the closest point to the origin is the same: $d_0$. The one parameter that identifies each sequence is $L_i$.

As we said before, one can consider that these two families correspond (formally) to a special choice of sharp test function. Let us see how the additive property holds in each family.

%\subsection*{First family type}

In the first one, it is enough to use invariance under time translation:

\begin{eqnarray}
S_4(\Theta(X_j)X_i)=S_4(-t^j_2, -t^j_1,t^i_1,t^i_2)=S_4(0, t^j_2-t^j_1,t^j_2 + t^i_1, t^j_2 + t^i_2)=\nonumber\\
S_4(0,L,t^j_1 + t^i_1+L, t^j_1 + t^i_1 + 2L)=F(t^{i}_1+t^{j}_1)\label{Seta}
\end{eqnarray}

In this case, the Schwinger functions appearing in the RP inequalities are functions of the sum of the starting points of each sequence, since $L$ is a fixed parameter. It is important to mention here that the first family has this property in any QFT, not just in CFT.

%\subsection*{Second family type}

For the second family, it is not immediate to identify the pd function of the sum of the parameters. In fact, in a general QFT there is not such function. However, we will see in the next section that in CFT the conformal invariance will allow us to rewrite the 4-point function $S_4(\Theta(X_j)X_i)$ as a function of the sum of the parameter $\log(\frac{2d_0}{L_i}+1)$.

 \section{RP of the 4 point function in CFT}\label{linkblock}

 In a general CFT the 4-point function $S_4(x_1,x_2,x_3,x_4)$ corresponding to 4 scalar fields of conformal dimensions $\Delta_i$ ($i=1...4$) can be written in terms of a function $g^{21}_{34}$ (that depends on the order of the fields) of two cross-ratio $u$ and $v$ defined by: $u=\frac{x^2_{12}x^2_{34}}{x^2_{13}x^2_{24}}$ and $v=\frac{x^2_{14}x^2_{23}}{x^2_{13}x^2_{24}}$. Here $x_{ij}$ is the distance between the points $x_i$ and $x_j$ in the $d$-dimensional Euclidean space.

\begin{equation}
S_4(x_1,x_2,x_3,x_4)=\left(\frac{x_{24}^{2}}{x_{14}^{2}}\right)^{\frac{1}{2} \Delta_{12}}\left(\frac{x_{14}^{2}}{x_{13}^{2}}\right)^{\frac{1}{2} \Delta_{34}} \frac{g^{21}_{34}(u, v)}{\left(x_{12}^{2}\right)^{\frac{1}{2}\left(\Delta_{1}+\Delta_{2}\right)}\left(x_{34}^{2}\right)^{\frac{1}{2}\left(\Delta_{3}+\Delta_{4}\right)}}, \label{relacion}
\end{equation}
where $\Delta_{ij}=\Delta_{i}-\Delta_{j}$.

As we mentioned in the previous section, we want to focus on the case in which the 4 points are aligned with $x_1<x_2<x_3<x_4$ and where in order to write the RP inequality of eq. (\ref{RP}) with $N=2$, we need to restrict to the case $\Delta_1=\Delta_4$ and $\Delta_2= \Delta_3$. This 4 points are identified with the $t^i_{\alpha}$ as: $x_1=-t^j_2$, $x_2=-t^j_1$, $x_3=t^i_1$, $x_4=t^i_2$. The Schwinger function (\ref{relacion}) can be written in terms of a single cross-ratio $x=\sqrt{u}$ (since in this case $v=(1-x)^2$) as follows:

 \begin{equation}
S_4(x_1,x_2,x_3,x_4)=\left(\frac{(1-x)^2}{x}\right)^{\Delta_{21}}(x^2_{23})^{-\Delta_{21}}(x^2_{12}x^2_{34})^{-\Delta_1}g^{21}_{21}(x)\label{mainlink}
\end{equation}

The positive definiteness requires then that: $\sum_{i,j}C_i {C^*}_jS_4(-t^j_2,-{t^j}_1,{t^i}_1,{t^i}_2)\geq{0}$.

Depending on the one parameter arrangement of points given by the $t_i$, this condition will be equivalent to the positive definiteness of a single variable function associated with each arrangement. Let us consider the two family of sequence introduced before 

\subsection[Positivity conditions in the $\rho-$family]{Positivity conditions in the $\boldsymbol{\rho-}$family}\label{rhoposcond}

In the $\rho-$family, it is not obvious that $S_4(\Theta(X_j)X_i)$ becomes a function of the sum of a suitable parameter of $X_j$ and $X_i$. In order to see that, let us define the parameter $\rho_i=\log(\frac{2d_0}{L_i}+1)$. The cross-ratio $x$ associated with the 4 points in the sequence $\Theta(X_j)X_i)$ becomes then $x=e^{-(\rho_i+\rho_j)}=e^{-\rho}$, with $\rho$ the sum variable. Evaluating (\ref{relacion}) in this particular case, we find:

\begin{equation}
\begin{aligned}
S_{4}(x_1,x_2,x_3,x_4)&=S_4(-t^j_2,-t^j_1,t^i_1,t^i_2)\\
&=\left(\frac{(1-x)^2}{x}\right)^{\Delta_{21}}(2d_0)^{-2\Delta_{21}}(L_i{L_j})^{-2\Delta_1}g^{21}_{21}(x)
\end{aligned}
\end{equation}

The positive definiteness requires that $\sum_{i,j}C_i {C^*}_jS_4(-t^j_2,-{t^j}_1,{t^i}_1,{t^i}_2)\geq{0}$. Since the $C_i$ are arbitrary numbers, and $x$ is the product $e^{-\rho_i}e^{-\rho_j}$, one can redefine
$$\tilde{C}_i=\left(\frac{e^{\rho_i}}{2d_0}\right)^{\Delta_{21}}{L_i}^{-2\Delta_1}C_i$$

and conclude that 
\begin{equation}\label{rhoposfunction}
(1-x(\rho))^{2\Delta_{21}}g^{21}_{21}(x(\rho))
\end{equation}

 is pd as a function of $\rho=-\log(x)$. Moreover, since it is bounded as $\rho\rightarrow{\infty}$, it has to be a CM function. This is equivalent to say that it should have a positive Laplace anti-transform and then the coefficients $b_n$ in the formal expansion:

\begin{equation}\label{RPpos}
(1-x)^{2\Delta_{21}}g^{21}_{21}(x)=\sum_{n} b_n x^n=\sum_{n} b_n e^{-n\rho}
 \end{equation}
 
should be all non-negative (where $n$ denotes a generic non-integer power). This coefficients are not identical to the ones of $g$, since there is a prefactor $(1-x)^{2\Delta_{21}}$. If we write formally $g^{21}_{21}(x)=\sum_m a_m x^m$ and insert the expansion:

$$(1-x)^{2\Delta_{21}}= \sum_{l=0}^{\infty}(-2\Delta_{21})_l \frac{x^l}{l!}$$

where $(-2\Delta_{21})_l$ is the Pochhammer symbol, the previous positivity conditions implies that the coefficients $b_n$:

\begin{equation}\label{bn}
b_n\equiv\sum_{l+m=n} (-2\Delta_{21})_l \frac{a_m}{l!} 
\end{equation}

must be non negatives.

Although these conditions seem different from the proven positivity of the coefficients of the expansion of $g$ itself mentioned previously in the literature (See Appendix A of \cite{Hartman:2015}), it can be seen that both are equivalent conditions. If we use the symmetry of the 4-point function under permutation $1\leftrightarrow2$ and $3\leftrightarrow4$, one can easily check that this implies the following identity:

\begin{equation}\label{gequality1234}
(1-x)^{2\Delta_{21}}g^{21}_{21}(x)=g^{12}_{12}(x)
\end{equation}

Using this equality one can see that $(1-x)^{2\Delta_{21}}g^{21}_{21}(x)$ is a CM function of $\rho$ for all values of $\Delta_1$ and $\Delta_2$ if and only if $g^{12}_{12}(x)$ is a CM function of $\rho$. And this last condition is equivalent to demanding that the coefficients ``$a$" appearing in the expansion of $g$ in powers of $x$ also must be non-negatives, in agreement with \cite{Hartman:2015}.

\subsection[Positivity conditions on the $\eta-$family]{Positivity conditions on the $\boldsymbol{\eta-}$family}

If we choose as the parameter defining the sequences $X_i$ in the $\eta-$family the normalized distance $\eta_i = t_i/L$, one can see that the cross-ratio $x$ is a function of the sum variable $\eta=\eta_i + \eta_j$: $x=\frac{1}{(1+\eta_i+\eta_j)^2}=\frac{1}{(1+\eta)^2}$. We can also check that this is the case for the other factors in (\ref{mainlink}) and it give us:

\begin{equation}
S_4(x_1,x_2,x_3,x_4)=L^{-2\Delta_1-2\Delta_2} \left(\frac{(1-x(\eta))^2}{x(\eta)}\right)^{\Delta_{21}}\eta^{-2\Delta_{21}}g^{21}_{21}(x(\eta)) \label{eta}
\end{equation}

that has to be a pd function of the variable $\eta$. Moreover, since this is bounded as $\eta\rightarrow\infty$, it implies that:
	
\begin{equation}
S_4(x_1,x_2,x_3,x_4)=L^{-2\Delta_1-2\Delta_2}\left(\frac{\eta+2}{\eta+1}\right)^{2\Delta_{21}}g^{21}_{21}\left(\frac{1}{(\eta+1)^2}\right),
\label{FuncionCasoEta}
\end{equation}
	
has to be completely monotonic (CM) as function of $\eta$. Note that (\ref{FuncionCasoEta}) differs from (\ref{rhoposfunction}) in the factors in front of $g$.

From the $\rho$ case in section \ref{rhoposcond} we see that $g^{21}_{21}(x)$ has a power expansion with positive coefficients of $x=\frac{1}{(1+\eta)^2}$. For any positive number $\alpha$, $\left(\frac{1}{(1+\eta)^2}\right)^{\alpha}$ is a CM function. Therefore as $g^{21}_{21}(\eta)$ is a combination of CM functions with positive coefficients. So when $\Delta_{21}=0$ the positivity conditions of the $\rho$ case imply that the conditions in the $\eta$ case are also fulfilled.

In the general case of $\Delta_{21}\neq 0$, we need to take into account the extra factor: $$\left(\frac{\eta+2}{\eta+1}\right)^{2\Delta_{21}}=\left(1+\sqrt{x(\eta)}\right)^{2\Delta_{21}}$$ that differs from the factor $(1-x)^{2\Delta_{21}}$ appearing in the positivity conditions (\ref{rhoposfunction}) of $\rho$. So in order to study the CM character of (\ref{FuncionCasoEta}), it is convenient to consider two separated case.

{\bf Case $\Delta_2>\Delta_1$}: Let us recall the form of the function (\ref{FuncionCasoEta}) which should be CM as a function of $\eta$ :

$$S_4(x_1,x_2,x_3,x_4)=\left(\frac{\eta+2}{\eta+1}\right)^{2\Delta_{21}}L^{-2\Delta_1-2\Delta_2}g^{21}_{21}(x(\eta))$$

Using the definitions and theorems of section \ref{pd}, one can see that $\left(\frac{\eta+2}{\eta+1}\right)^{2\Delta_{21}}$ is a CM function. This can easily be done by showing that its logarithm (when $\Delta_{21}>0$) is a CM function and so it will also be its exponential.

Therefore, $S_4$ is a CM function of $\eta$, as it is a sum with positive coefficients in which each term is a product of CM functions, $\left(\tfrac{\eta+2}{\eta+1}\right)^{2\Delta_{21}} \left(\frac{1}{(1+\eta)^2}\right)^{\alpha}$.

{\bf Case $\Delta_2<\Delta_1$}: This case is more involved, since in this case $\left(\frac{\eta+2}{\eta+1}\right)^{2\Delta_{21}}$ is not a complete monotonic function. So, the CM character of $S_4$ is not guaranteed by the positivity of coefficients in $g$, meaning that in this case we have extra conditions. The CM character of $S_4(\eta)$ means that:

$$(-1)^n S_4^{(n)}(\eta)\geq{0},$$

or equivalently, that its Laplace anti-transform is non-negative.

\subsection{ Checking positivity of coefficients in conformal blocks}

In this subsection, we want to see how the previous considered positivity conditions are fulfilled in the conformal block itself. We will focus in the particular case of 2d CFT, considering all the Virasoro descendants. In this 2 dimensional case the check can be done in an elegant way and we use it to illustrate the idea, but this positivity checks can be done, at least numerically, for conformal blocks in general dimensions.

Now, we use the conformal block decomposition of $g$:
\begin{equation}
	g_{34}^{21}(x, \bar{x})=\sum_{p} C_{34}^{p} C_{12}^{p} \mathcal{F}_{34}^{21}(p | x) \bar{\mathcal{F}}_{34}^{21}(p | \bar{x}),\label{expanciondeg}
\end{equation}
where
\begin{equation}\label{expansiondebloques}
\begin{aligned}
&\mathcal{F}_{34}^{21}(p | x)=x^{h_{p}} \sum_{k=0}^{+\infty} F^{(p)}_{k} x^{k}\\
&\bar{\mathcal{F}}_{34}^{21}(p | \bar{x})=\bar{x}^{\bar{h}_{p}} \sum_{k=0}^{+\infty} \bar{F}^{(p)}_{k} \bar{x}^{k},
\end{aligned}
\end{equation}
where $x$ and $\bar{x}$ are the holomorphic and antiholomorphic cross-ratios, that are equal when the points are aligned and ordered, and $\Delta_p=h_p+\bar{h}_p$, $\ell_p=h_p-\bar{h}_p$ are the conformal dimension and spin of the exchange fields.

RP inequalities applies in the case in which $1=4$ and $2=3$. In this case, we have:

\begin{equation}
	g_{21}^{21}(x, \bar{x})=\sum_{p} ( C_{12}^{p})^2 \mathcal{F}_{21}^{21}(p | x) \bar{\mathcal{F}}_{21}^{21}(p | \bar{x}),\label{expanciondeg}
\end{equation}

As we have mentioned in the introduction, we known from \cite{RychkovRP} that RP is automatically fulfilled in the 4-point functions, without restricting the OPE coefficients or the field dimensions (besides unitary bound and reality of the OPE coefficient). Then, from the simple fact that the $C$'s appears as $C^2$ and they are real, from the positivity of the coefficients of $g$ we conclude that the conformal blocks should also have an expansion with positive coefficients, because if not, there will be further restrictions on the CFT data\footnote{Actually, what appears is $C_{21}^{p}C_{12}^{p}=(-1)^{\ell_p} C_{12}^2$, where $\ell_p$ is the spin of the exchange field. The minus sign for odd spin cancels with other minus from the conformal blocks.}. This can be checked explicitly using the exact form of the conformal blocks. 

\subsubsection[$\rho-$family]{$\boldsymbol{\rho-}$family}\label{rhocheck}

Let us consider the implications of RP for the $\rho-$family and let us check this in the two dimensional case with invariance under local conformal transformations where we have the closed form of the coefficients. In this case the function $\mathcal{F}$ admits the following form:
\begin{equation}{\label{bloques}}
	\mathcal{F}\left(c, h_{i}, h_{p}, x\right)=\sum_{q=0}^{\infty} x^{h_{p}+q} \chi_{q}\left(c, h_{i}, h_{p}\right)\ { }_{2}F_{1}\left[\begin{array}{c}
	h_{p}+q+h_{21},\ h_{p}+q+h_{21} \\
	2\left(h_{p}+q\right)
	\end{array} ; x\right]
\end{equation}

where the $q-$sum is over the contribution of the descendants (see for instance, eq (2.11) of \cite{Perlmutter:2015iya}), and the antiholomorphic function $\mathcal{\bar{F}}$ admits an analogous expression.

The coefficients $\chi_{q}\left(c, h_{i}, h_{p}\right)$ are algebraic expressions completely determined by the conformal symmetry.

Because of the form of hypergeometric functions

\begin{equation}\label{2F1}
{ }_{2}F_{1}\left[\begin{array}{c}
a,\ b \\
c
\end{array} ; x\right]=\sum_{n=0}^{\infty} \frac{(a)_{n}(b)_{n}}{(c)_{n}} \frac{x^{n}}{n !}
\end{equation}
we see that the coefficients of of the hypergeometric functions are positive since $a=b=h_{p}+q+h_{21}=$real, and so $(a)_n (b)_n=(a)^2_n \geq{0}$, together with $c$ being positive because $h$ and $q$ are positive, the coefficients of the ${}_2F_1$ expansion are positive. The same happens with the antiholomorphic version of (\ref{bloques}). 
As $\chi$ is positive, then the coefficients in the power expansions of $g$ are positive too. Then by (\ref{gequality1234}) equality, also the $b_n$ coefficients in (\ref{RPpos}) are positive and the positivity condition coming from RP holds.

In fact, the positivity of the $b_n$ coefficients can also be checked directly from the expression
$$(1-x)^{2(\Delta_2-\Delta_1)}g^{21}_{21}(x)$$
and some properties of the hypergeometric functions ${}_2F_1$. While $g$ has non-negative coefficients, the first factor $(1-x)^{2(\Delta_2-\Delta_1)}$ contains terms with negative coefficients if $\Delta_2>\Delta_1$. This leads to the possibility that some $b_n$ coefficients in (\ref{bn}) may be negative, and then have more restrictions comparing to \cite{Hartman:2015}. As we have seen, this is not the case. To check that all $b_n$ are indeed positive it will be useful to decompose
\begin{equation}
	(1-x)^{2(\Delta_2-\Delta_1)}=(1-x)^{2(h_2-h_1)} (1-x)^{2(\bar{h}_2-\bar{h}_1)}
\end{equation}

Let us see the holomorphic part of eq (\ref{expanciondeg}) using eq (\ref{expansiondebloques}): multiplying $(1-x)^{2(h_2-h_1)} \mathcal{{F}}$ inside the sum: 
\begin{equation}{\label{hyper}}
	\begin{aligned}
		(1-x)^{2(h_2-h_1)} &{ }_{2}F_{1}\left[\begin{array}{c}
		h_{p}+q+h_{21},\ h_{p}+q+h_{21} \\
		2\left(h_{p}+q\right)
		\end{array} ; x\right]
		\\
		\\ =&{ }_{2}F_{1}\left[\begin{array}{c}
		h_{p}+q-h_{21},\ h_{p}+q-h_{21} \\
		2\left(h_{p}+q\right)
		\end{array} ; x\right]
	\end{aligned}
\end{equation}
where in the last equality we use an hypergeometric identity. Because of the form of the hypergeometric function (\ref{2F1}), again we see that the coefficients are positive, since $a=b=h_{p}+q-h_{21}=$ real, and so $(a)_n (b)_n=(a)^2_n \geq{0}$, together with $c$ being positive because $h$ and $q$ are positive, the coefficients of the ${}_2F_1$ expansion are positive. The same happen with the antiholomorphic part.

Then, despite the potentially negative contribution of the prefactor $(1-x)^{2(h_2-h_1)}$ the coefficients in $(1-x)^{2(h_2-h_1)} \mathcal{{F}}$ turn to be non-negative. The final step in the proof is to check that the factors $\chi_{q}\left(c, h_{i}, h_{p}\right)$ does not spoil this positivity. It is easy to see if we look at the equation 2.12 of \cite{Perlmutter:2015iya}.

So, the Virasoro conformal blocks are fixed by symmetry and we saw that these stronger positivity conditions are already fulfilled in each conformal block separately. Since $g$ is a combination of conformal block weighted by positive coefficients (the square of the OPE coefficients) the RP restricted to this family of sequences are automatically fulfilled without imposing any further restrictions to the OPE coefficients beyond unitary bound.\footnote{Similarly, in the $d$ dimension using the closed form of the conformal blocks \cite{Diagonal}, one can check also that the positive conditions are fulfilled.}

\subsubsection[$\eta-$family]{$\boldsymbol{\eta-}$family}

As we said before, the only non-trivial case is $\Delta_{21}>0$. This case can be checked explicitly in several examples. The positivity condition can be state in the form

\begin{equation}
\left(1+\sqrt{x(\eta)}\right)^{2\Delta_{21}}g^{21}_{21}(x(\eta))=\int_{0}^{+\infty} F(s)e^{-s\eta}ds
\end{equation}

with $F(s)\geq{0}$ and as before $x(\eta)=\frac{1}{(\eta+1)^2}$.

This is a less obvious example of a positivity condition implied by RP, different from the positivity condition of the coefficients of $g$ showed above in section \ref{rhocheck} and previously in \cite{Hartman:2015}.

\section{Combining crossing symmetry with reflection positivity}\label{CrossinPositivity}

\subsection[Warming up with 2$d$ Ising Model]{Warming up with 2$\boldsymbol{d}$ Ising Model}
The strategy can be understood in the following pedagogical example. Let us consider the Ising model in $d=2$, which is a minimal model corresponding to $c=\frac{1}{2}$ with three spinless primary fields of total conformal dimension $\Delta=h+\bar{h}=0,1/8,1$ ($h=\bar{h}$). As usual, let us denote these fields with $1,\sigma,\epsilon$ respectively. The OPE coefficients are partially determined by the fusion rules, which says in particular that:

$$[\sigma\sigma]=[1] +[\epsilon]$$

It means that the OPE coefficients $C_{\sigma\sigma}^{[\sigma]}=0$.
$C_{\sigma\sigma}^{[1]}$ can be normalized to 1 while $C_{\sigma\sigma}^{[\epsilon]}$ ($C$ from now on) is not fixed by the fusion rules alone. However, the value of $C$ is completely fixed by CS, which impose that:

\begin{eqnarray}
G^{\sigma\sigma}_{\sigma\sigma}(x) &=& G^{\sigma\sigma}_{\sigma\sigma}(1-x) \ \text{\footnotemark} \rightarrow\nonumber\\
x^{-2\Delta_{\sigma}}(F^{\sigma\sigma}_{\sigma\sigma}[1](x)+C^2F^{\sigma\sigma}_{\sigma\sigma}[\epsilon](x))&=&
(1-x)^{-2\Delta_{\sigma}}(F^{\sigma\sigma}_{\sigma\sigma}[1](1-x)+C^2F^{\sigma\sigma}_{\sigma\sigma}[\epsilon](1-x))\nonumber
\end{eqnarray}
\footnotetext{In general, G is related to g, by $G^{21}_{34}(x)=x^{-(\Delta_{3}+\Delta_{4})} g^{21}_{34}(x)$.}
The $F^{\sigma\sigma}_{\sigma\sigma}[1]$ and $F^{\sigma\sigma}_{\sigma\sigma}[\epsilon]$ are the conformal blocks corresponding to the exchange fields $1$ and $\epsilon$, which are completely fixed by general properties of the CFT:

\begin{equation}
\begin{aligned}
F[1](x)=\frac{1}{(1-x)^{\frac{1}{4}}}(1+\sqrt{1-x})\\
F[\epsilon](x)=\frac{1}{(1-x)^{\frac{1}{4}}}(1-\sqrt{1-x})
\end{aligned}
\end{equation}

where we have suppressed the index $\sigma$.

Witting down the rhs of te CS equation:

\begin{equation}
rhs=\frac{1}{(1-x)^{\frac{1}{4}}}(1+C^2+(1-C^2)\sqrt{x})
\end{equation}
 one can see that, in order to get a CM function of $\rho$ in the rhs, the coefficient of $\sqrt{x}$ must be not negative: $C^2\leq{1}$.

 This example illustrates the main idea of this paper: the lhs is automatically CM, since the conformal blocks itself satisfies this conditions and they are combined with non-negative coefficients. Then the rhs should be a CM function of $\rho$ as well. But this is not guaranteed by the form of the conformal blocks. In fact, the first term is a CM function of $\rho$ but the second term is not. This imposes an upper bound to the non-CM term, which in this case turns to be $C^2\leq{1}$. Let us formulate the procedure in the more complicated case where we have an infinite number of OPE coefficients.

\subsection[General procedure in the ${d}$-dimensional CFT]{General procedure in the $\boldsymbol{d}$-dimensional CFT}\label{generalproc}

Let us write again the crossing symmetry equation for four identical scalar fields of conformal dimension $\Delta_0$, but now isolating the contribution of the identity at both sides:

\begin{equation}
	\frac{1}{x^{2\Delta_0}}\left[1+\sum_{\Delta, \ell} C^2_{\Delta, \ell}\ F^{\Delta, \ell}(x)\right]=	\frac{1}{(1-x)^{2\Delta_0}}\left[1+\sum_{\Delta, \ell} C^2_{\Delta, \ell}\ F^{\Delta, \ell}(1-x)\right]
\end{equation}
where the sum is over all conformal dimension $\Delta$ (except the identity $\Delta=0$) and spin $\ell$ of the exchange fields, and we have simplify the notation in the $C$'s suppressing the external field label $\Delta_0$.

We can see that in the rhs we have the conformal block corresponding to the identity, i.e., the term $\frac{1}{(1-x)^{2\Delta_0}}$. This term is already a CM function and therefore does not help in our strategy of bounding the OPE coefficients demanding the rhs to be CM. As adding a pd function would lead to less constrains, it is better not to have pd functions in the rhs. So, to get a CM lhs, and a rhs without the contribution of the identity, our strategy will be to rewrite the equation in the following way:

\begin{equation}
	1 + \frac{1}{1-(\frac{x}{1-x})^{2\Delta_0}}\sum_{\Delta, \ell} C^2_{\Delta, \ell}\ F^{\Delta, \ell}(x)=	\frac{x^{2\Delta_0}}{(1-x)^{2\Delta_0}-x^{2\Delta_0}} \sum_{\Delta, \ell} C^2_{\Delta, \ell}\ F^{\Delta, \ell}(1-x)
\label{laposta}
\end{equation}

In the lhs we have a sum of CM functions of $\rho=-\log(x)$ for $\rho>\log(2)$, this can be checked by noticing that the factor 
$$\frac{1}{1-(\frac{x}{1-x})^{2\Delta_0}}=\frac{1}{1-(\frac{e^{-\rho}}{1-e^{-\rho}})^{2\Delta_0}}$$
is a CM function of $\rho$ in this region. 

Now, we are ready to formulate our strategy in a precise way demanding that the rhs of eq (\ref{laposta}) must be a CM function of $\rho$ \footnote{No new restriction will come from the complete monotonicity of $\eta$ because, as said before, CM in $\rho$ imply CM in $\eta$ when all four fields are identical.}. If we call:
$$\tilde{F}_{\Delta, \ell}(\rho)\equiv	\frac{x^{2\Delta_0}}{(1-x)^{2\Delta_0}-x^{2\Delta_0}}  F^{\Delta, \ell}(1-x), \ \ \ \text{with}\ x=e^{-\rho}$$
to each contribution in the rhs of eq. (\ref{laposta}), one can formulate that the entire rhs could be CM in three equivalents ways:

\begin{enumerate}
\item {\bf Linear and local inequalities} The formal sum of the $n$ order derivatives of each contribution should be non-negative: $(-1)^{n}\sum_{\Delta, \ell} C^2_{\Delta, \ell}\ \tilde{F}^{(n)}_{\Delta, \ell}(\rho)\geq{0}$
\item {\bf Non Linear local inequalities} : The determinants of order $N$, equations (\ref{det1}, \ref{det2}), of the complete rhs, should be non-negative. This inequalities should hold for every value of $\rho \in$ $(\log(2),+\infty)$
\item {\bf Linear but global analysis} The Laplace anti-transform must be non-negative, i.e. the expansion in $e^{-\lambda \rho}$ (for $\lambda >0$) should have non-negative coefficients.
\end{enumerate}

\subsection{Linear analysis of the rhs}\label{linearRHS}

The conformal blocks corresponding to external spinless fields in $d$-dimensions, for 
the case relevant for us in which the 4 points are aligned were found in closed form in \cite{Diagonal}.

Using these results, let us consider first the conformal block corresponding to spinless exchange fields:

$$\left(\frac{x^2}{(1-x)}\right)^{\Delta /2} \, { }_{3}F_{2}\left[\begin{array}{l}
\frac{\Delta}{2}, \frac{\Delta}{2}, \frac{\Delta}{2}-\Delta _{\min} \\
\frac{\Delta+1}{2}, \Delta -\Delta_{\min}
\end{array} ; -\frac{x^2}{4(1-x) }\right]$$

where $\Delta_{min}=\frac{d-2}{2}$ corresponds to the minimal conformal dimension of a spinless field.

According with the previous discussion, the contributions of spinless fields in the rhs of the rewritten CS is a sum, weighted by non-negative coefficients, of the following function:

\begin{equation}
\tilde{F}_{\Delta, 0}(\rho)=\frac{{x}^{\Delta_0}}{1-(\frac{x}{1-x})^{2\Delta_0}}\left(\frac{(1-x)^2}{x}\right)^{\Delta /2-\Delta_0} \, { }_{3}F_{2}\left[\begin{array}{l}
\frac{\Delta}{2}, \frac{\Delta}{2}, \frac{\Delta}{2}-\Delta _{\min} \\
\frac{\Delta+1}{2}, \Delta -\Delta_{\min}
\end{array} ; -\frac{(1-x)^2}{4 x}\right]\nonumber
\end{equation}

The necessary condition for the rhs to be a solution of the CS equation is:

\begin{equation}
(-1)^n\sum_{\Delta}C^2_{\Delta} \tilde{F}^{(n)}_{\Delta, \ell}(\rho)\geq{0}\label{conditionH}
\end{equation}

being $\tilde{F}^{(n)}$ de $n-$derivative respect to $\rho=-\log(x)$. Let us ignore for the moment the contribution of all the other spins in the rhs, which in this case are only even \footnote{That is because one can see that, for spinless fields $i,j$, the OPE coefficients obey $C_{ij}^k=(-1)^{\ell_k}C_{ji}$, where $\ell_k$ denote the spin of the exchange field $k$. In this case, $i=j$, so the OPE coefficient vanishes when $\ell_k$ is odd. See for example, equation 2.2.48 in \cite{Ribault}.}.

This is a family of (formal) inequalities, which impose restriction on the $C$'s since most of the terms individually does not fulfills the inequalities. 

\begin{figure}[h]
	\centering
	\includegraphics[scale=.60]{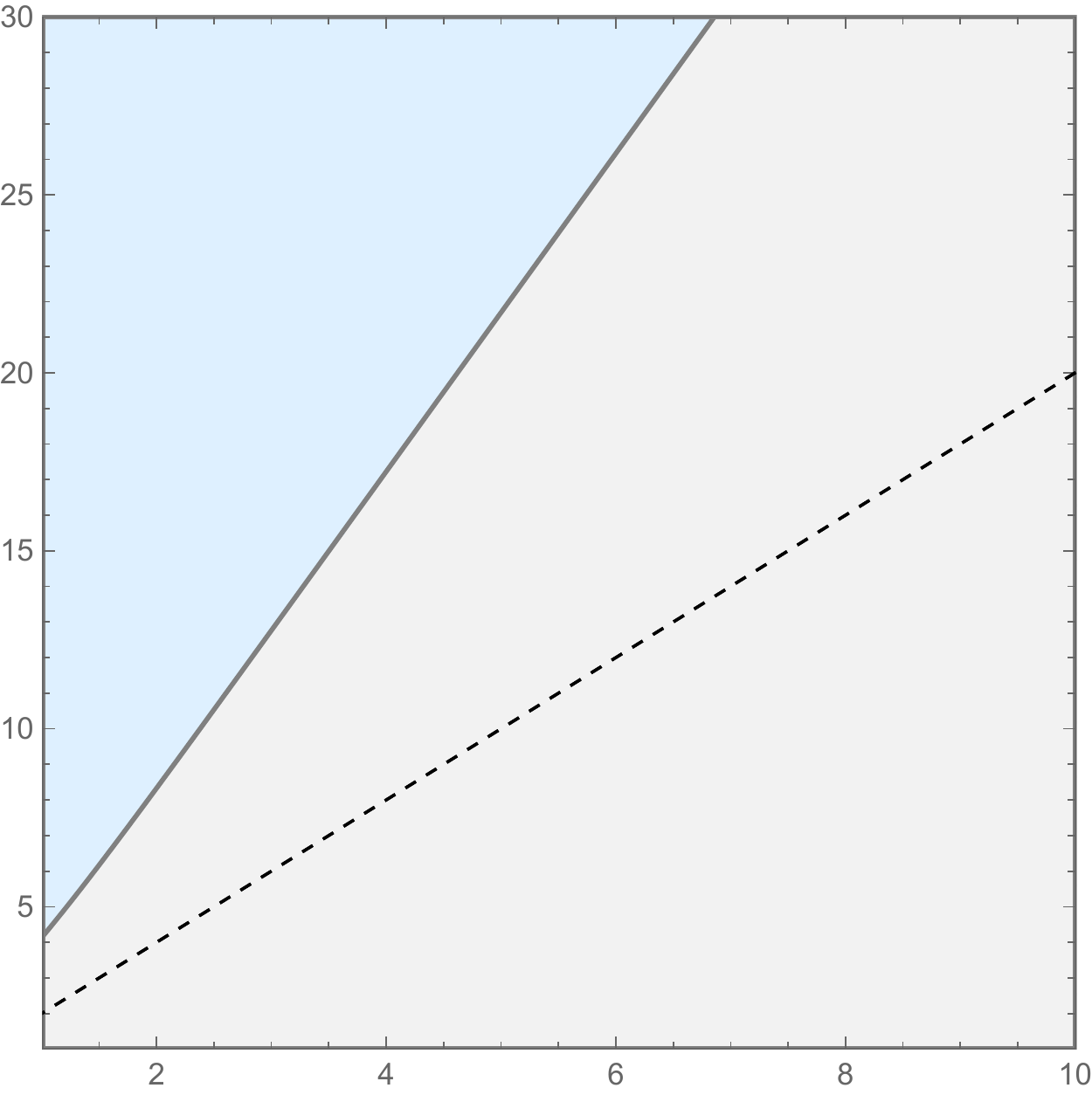}
	\caption{The blue region is the set of pairs of conformal dimensions $(\Delta_0,\Delta)$ such that the contribution of a spinless exchange field in the rhs fails in its first derivative to be a CM function. $\Delta_0$ is the dimension of the external field in the 4 point function of 4 identical scalar fields. Since the entire rhs should a a pd function, it is required the presence of exchange fields in the gray area. The dotted line correspond to $\Delta=2\Delta_0$ was given just as guide for the eye.} \label{Fig:newbound4d}
\end{figure}

Let us consider for example the case $d=4$, looking only at spinless exchange fields just to understand the idea. In Figure \ref{Fig:newbound4d} we have displayed the area in the plane $\{\Delta_0,\Delta\}$ where the first derivative of $\tilde{F}$ is negative(grey) or positive(blue). So, for $n=1$, as the complete sum in (\ref{conditionH}) must be positive, and because $C^2_{\Delta}>0$, necessarily at least one of the exchange field in the grey region must have a non-vanishing structure constant associated to $\Delta_0$. It means that the minimum $\Delta$ value of the exchange fields is bounded by the border line of the blue-gray regions, because only in the gray region the first derivative of $\tilde{F}$ has the right sign (negative) for a CM function.

\begin{figure}
	\centering
	
	$\begin{array}{cc}
	
	\begin{subfigure}[b]{0.4\textwidth}
		\subfloat[$\ell=2$]
		{\includegraphics[width=\textwidth]{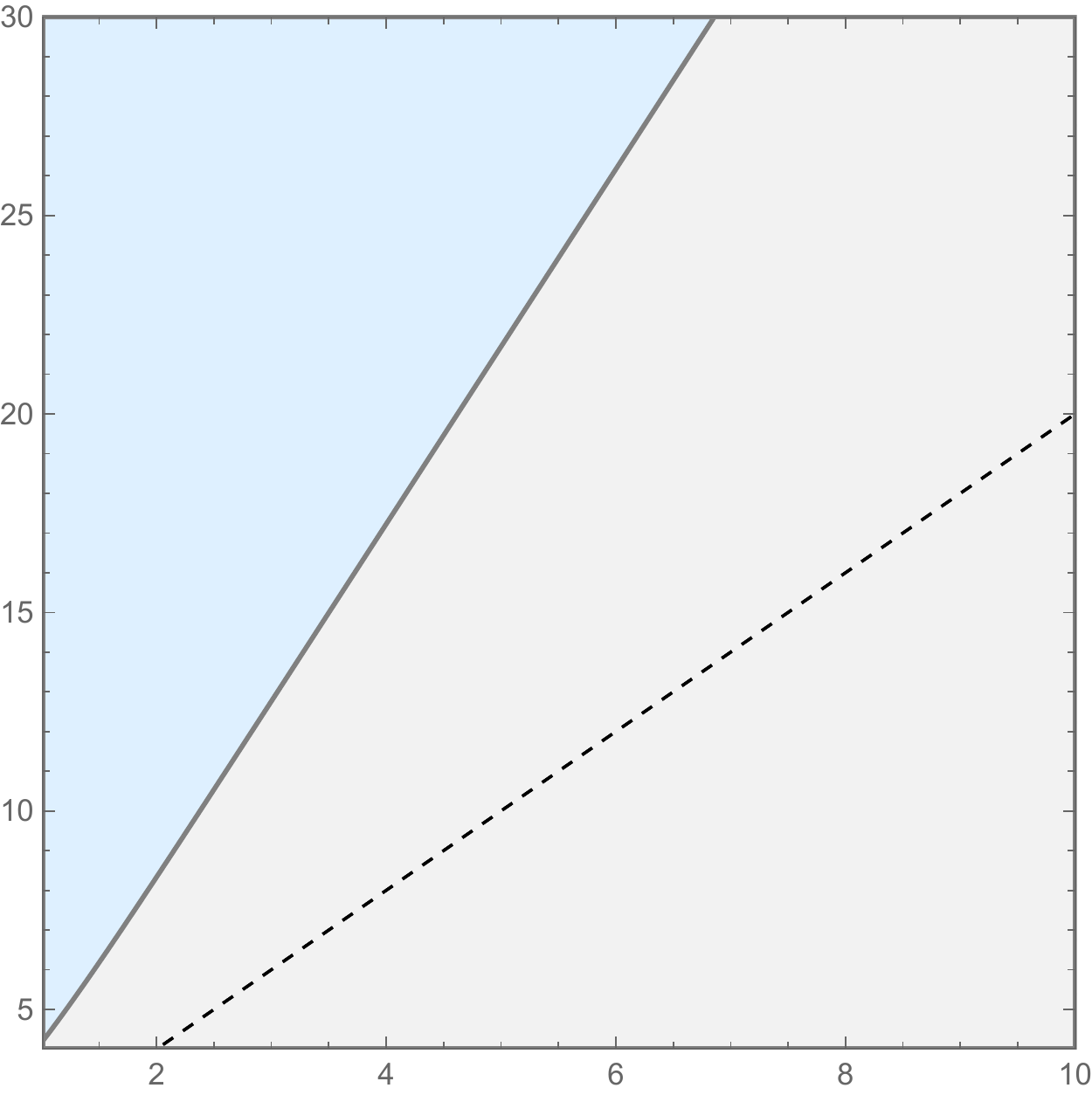}}
	\end{subfigure}
	&
	\begin{subfigure}[b]{0.4\textwidth}
		\subfloat[$\ell=4$]
		{\includegraphics[width=\textwidth]{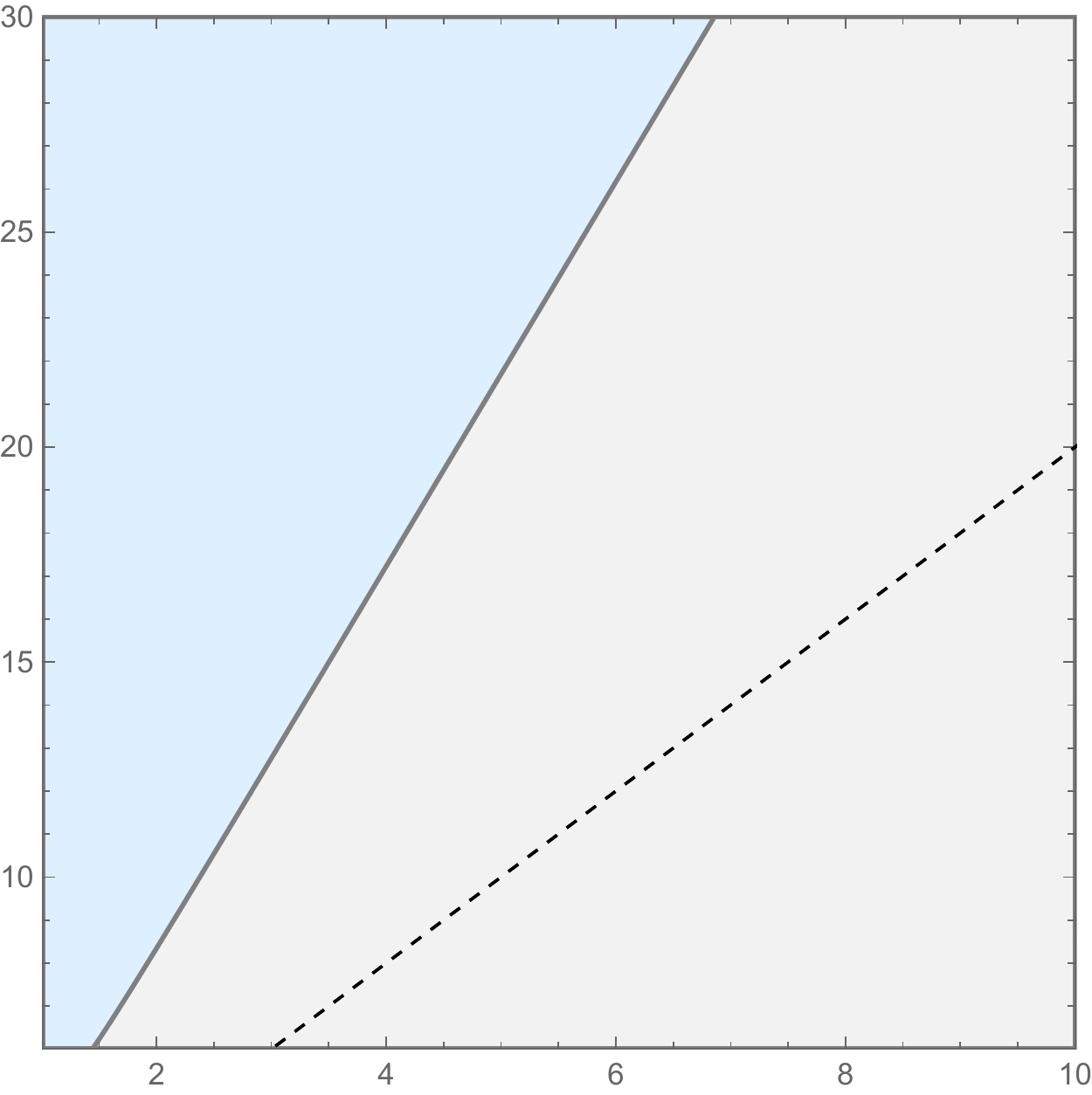}}
	\end{subfigure}
	
	\\
	
	\begin{subfigure}[b]{0.4\textwidth}
		\subfloat[$\ell=6$]
		{\includegraphics[width=\textwidth]{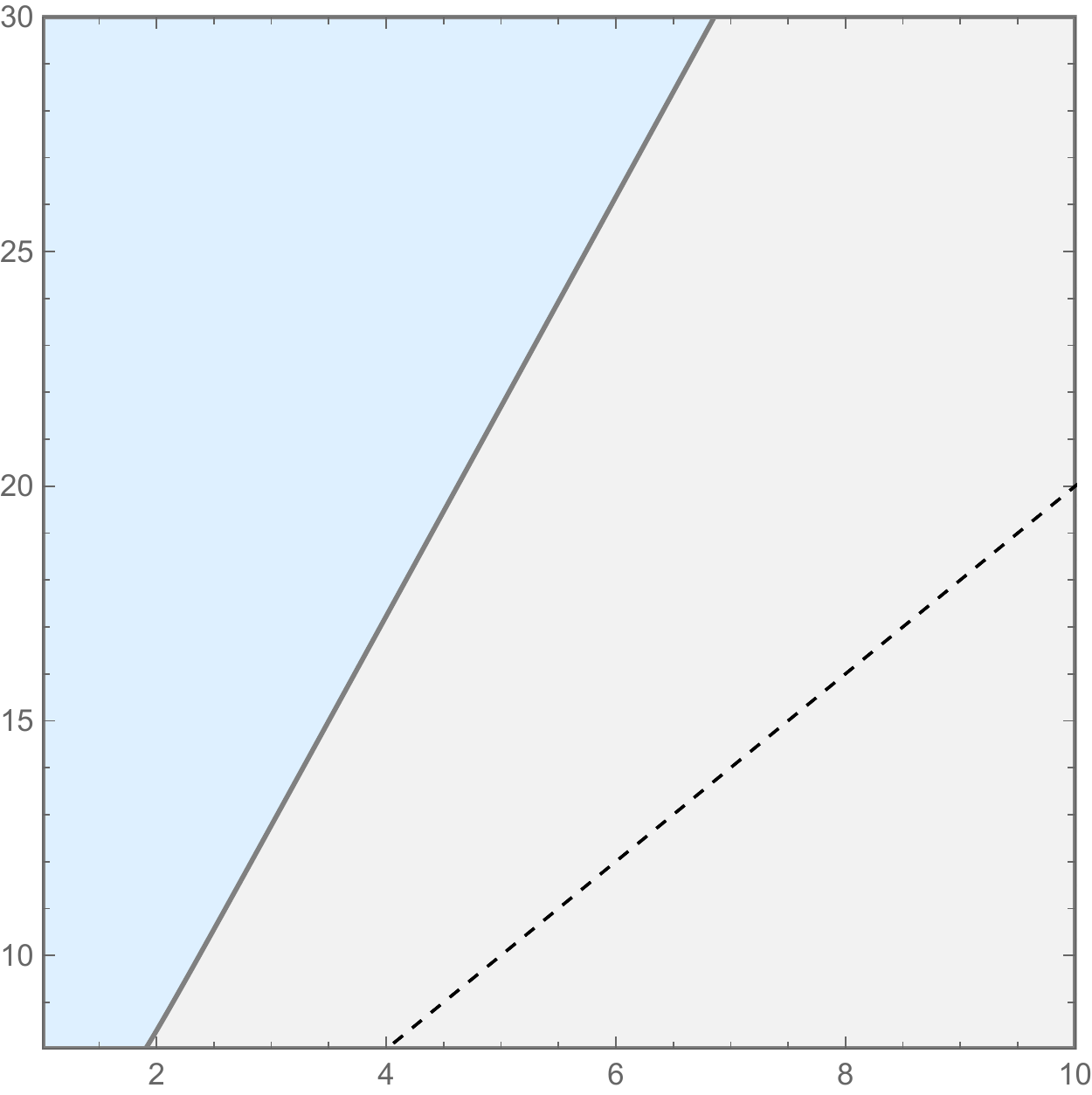}}
	\end{subfigure}
	&
	\begin{subfigure}[b]{0.4\textwidth}
		\subfloat[$\ell=8$]
		{\includegraphics[width=\textwidth]{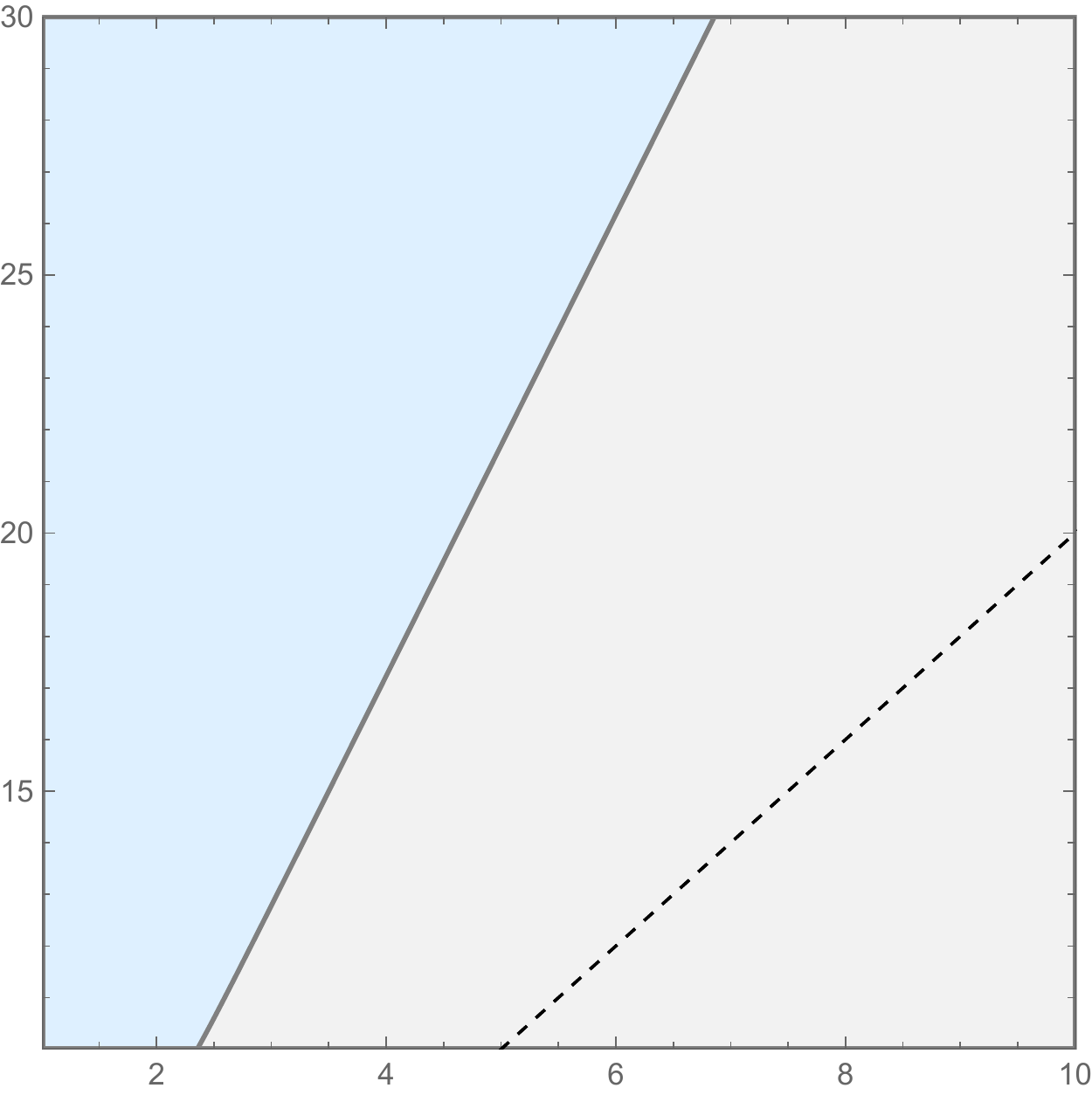}}
	\end{subfigure}

\end{array}$

\caption{Similar regions as in Figure \ref{Fig:newbound4d}, but for exchange fields of higher spins.}

\label{Fig:newbound4d_spins}
\end{figure}

Actually, one needs to include the contributions coming from all exchange fields with even spins. For a given $\Delta_0$, it is needed a $\Delta$ value in the gray region of \textit{any} of the plots like those in Figure \ref{Fig:newbound4d_spins}, where we have shown only the first spin contributions.
For $\Delta_0$ close enough to $1$, the previous conclusion can be strengthened. Looking for instance to values of $\Delta_0$ in $(1,1.4)$, one can conclude that {\it it should appear an exchange field contribution of spin $\ell=0$ or $\ell=2$, with its conformal dimension in the gray area}. This can be seen by checking that in higher spin plots the border blue-gray line moves to the right as the spin increase, as it happens from $\ell=4$ to $\ell=8$ in the Figure. 

In order to turn this numerical argument into a rigorous one, we need to prove that all the infinite higher spin conformal blocks contribute with a derivative of wrong sign. We don't have an analytical proof of this, but we checked it numerically for high values of even spin (from $\ell=2$ to $\ell=50$). A simple way to be convinced of this, is that as $\ell \geq 4$ increases, the region of $\Delta_0$ with no $\Delta$ with correct (negative) first derivative sign also increases. For $\ell=0,2$ for any $\Delta_0$ there is always a $\Delta$ with the correct first derivative sign, but for higher spins, considering unitary bound ($\Delta \geq \ell+d-2$), there are regions of $\Delta_0$ that don't have any $\Delta$ with correct first derivative sign. For example looking at the plots in Figure \ref{Fig:newbound4d_spins}, this region for $\l=4$ is $\Delta_0 \in (1,1.4)$, for $\l=6$ is $(1, 1.9)$ and for for $\l=8$ is $(1,  2.4)$, approximately. This tendency is observed for all the plots of higher spin values evaluated. Nevertheless, for example, one can also state, that ``for $\Delta_0$ approximately in $(1.4,1.9)$ at least one exchange field contribution in the greys areas of spin $\ell=0$ or $\ell=2$ or $\ell=4$ should be present", this is because in this region, for $\ell \geq 6$ there is no $\Delta$ with the correct first derivative sign, but there is for $\ell=0, 2, 4$. One can make similar statements for other $\Delta_0$ regions.

\begin{figure}
	\centering
	
	$\begin{array}{cc}
	
	\begin{subfigure}[b]{0.4\textwidth}
		\subfloat[$2$nd derivative]
		{\includegraphics[width=\textwidth]{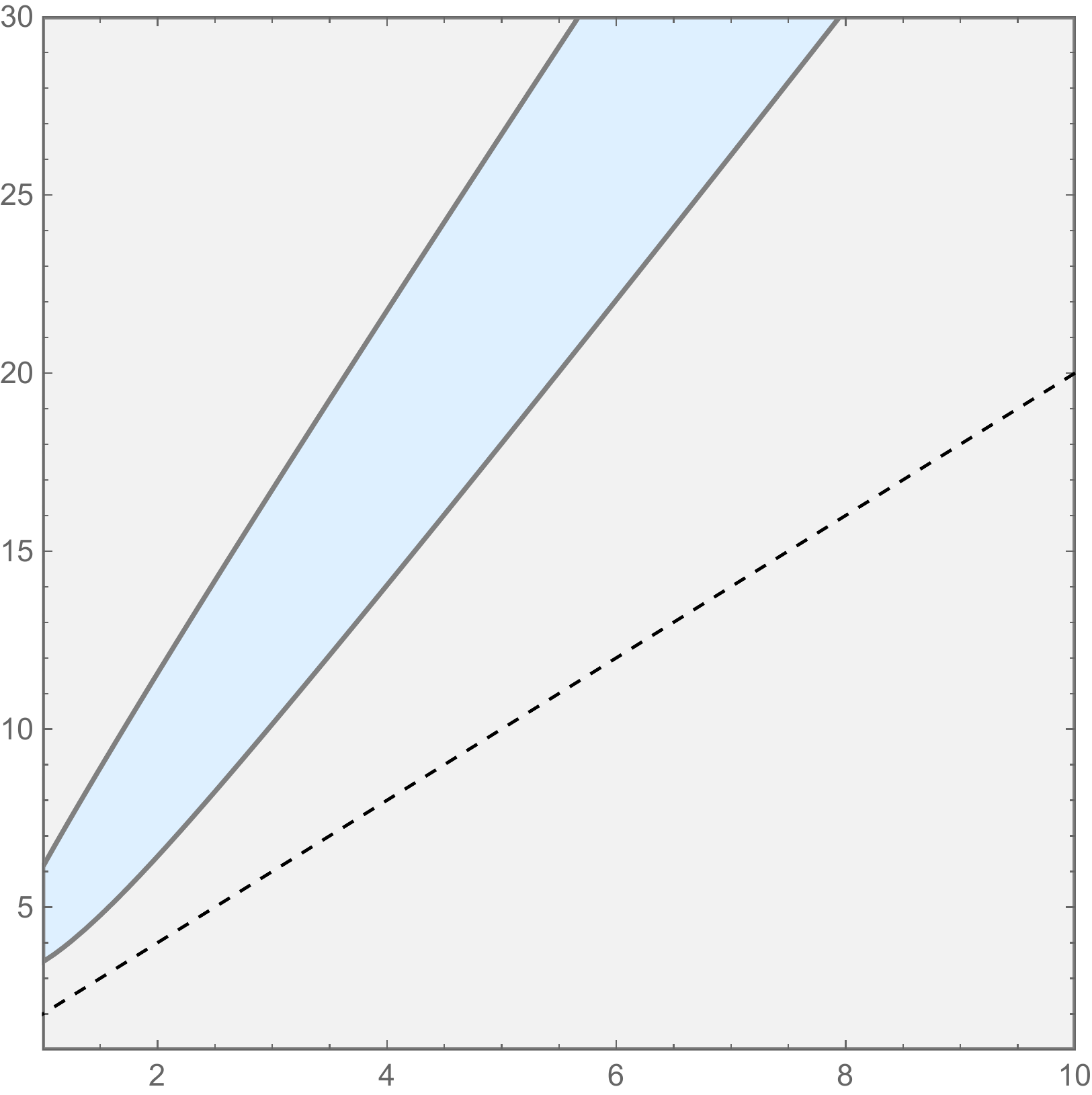}}
	\end{subfigure}
	&
	\begin{subfigure}[b]{0.4\textwidth}
		\subfloat[$3$rd derivative]
		{\includegraphics[width=\textwidth]{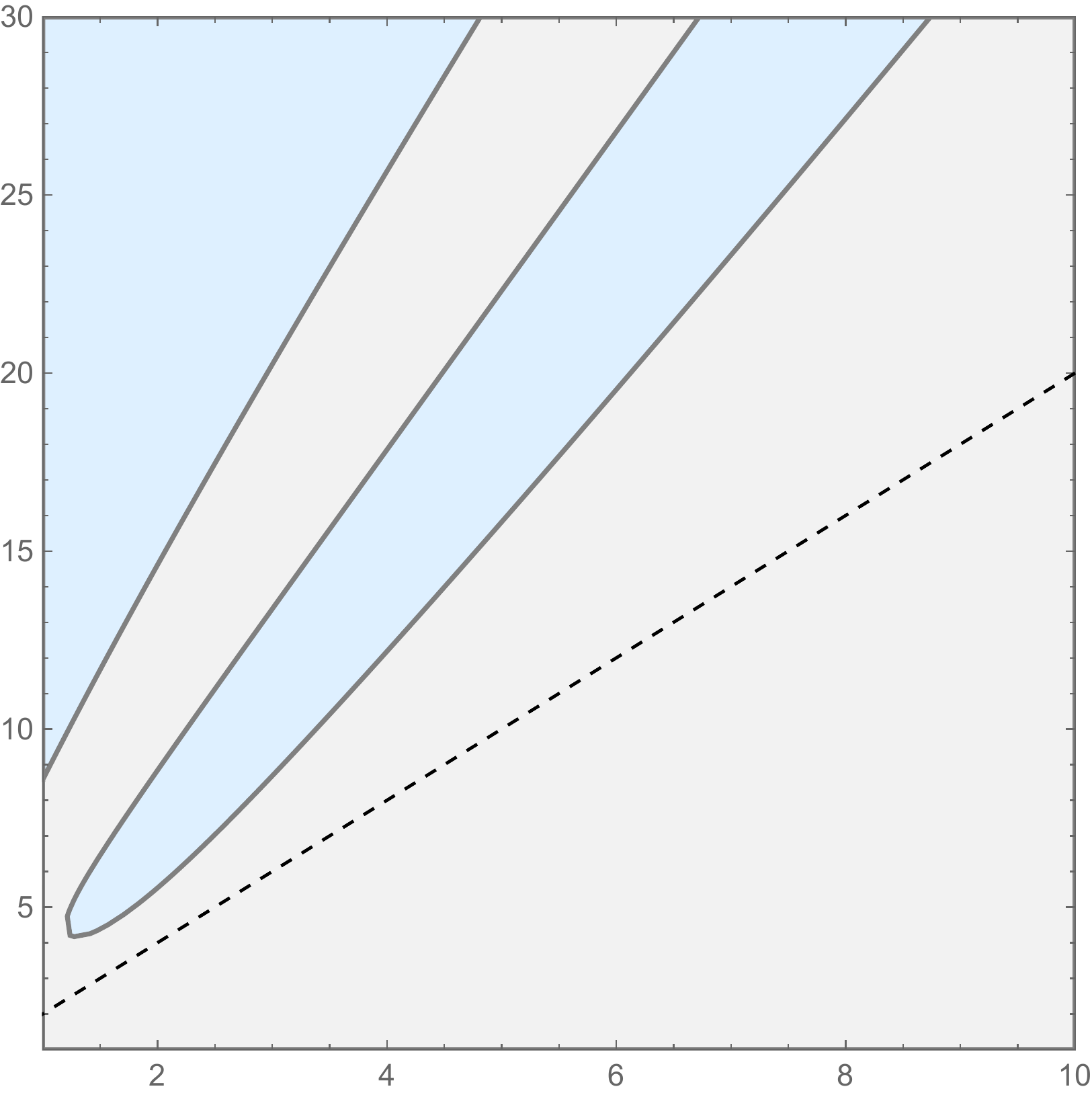}}
	\end{subfigure}
	
	\\
	
	\begin{subfigure}[b]{0.4\textwidth}
		\subfloat[$4$th derivative]
		{\includegraphics[width=\textwidth]{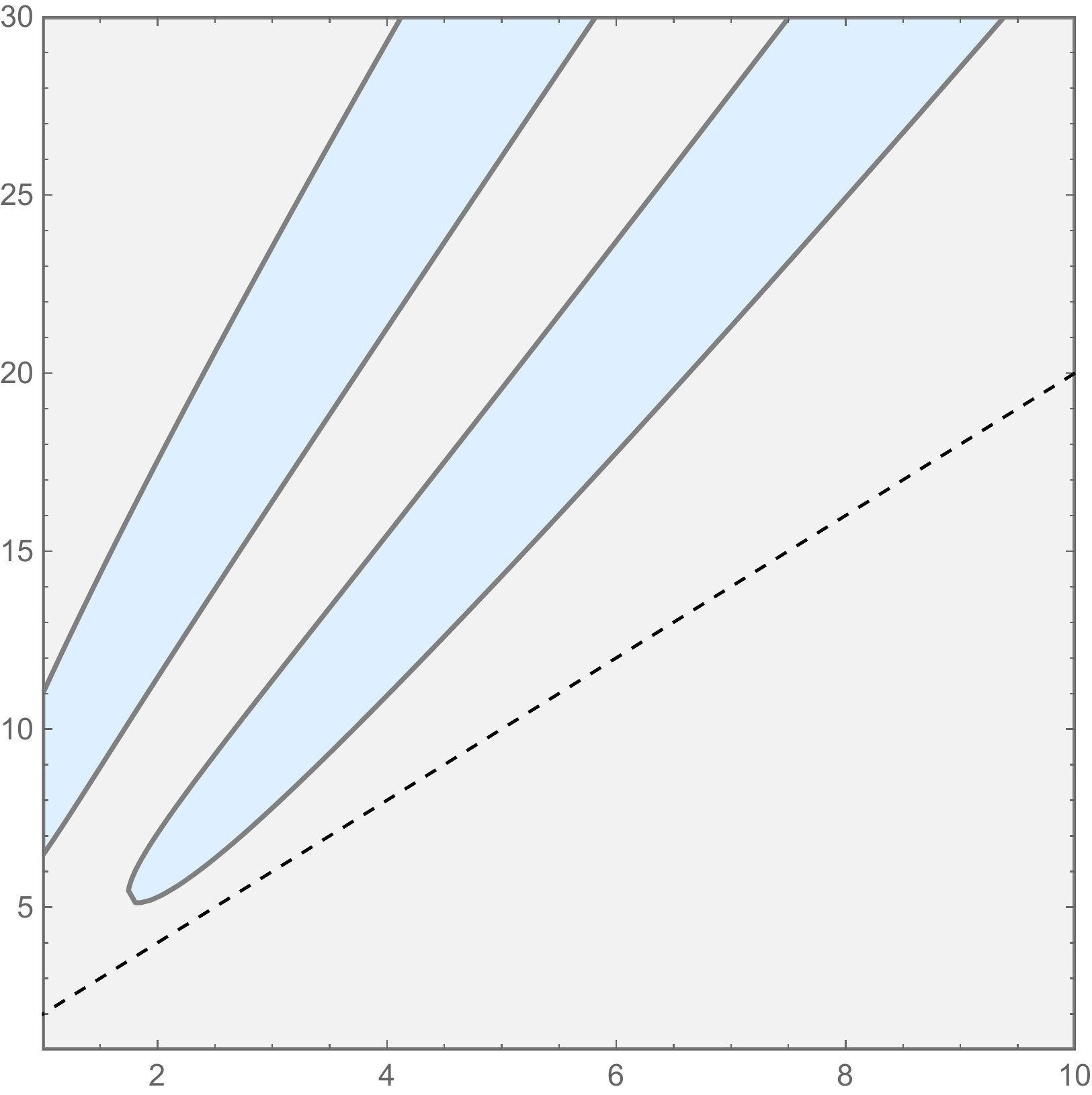}}
	\end{subfigure}
	&
	\begin{subfigure}[b]{0.4\textwidth}
		\subfloat[$5$th derivative]
		{\includegraphics[width=\textwidth]{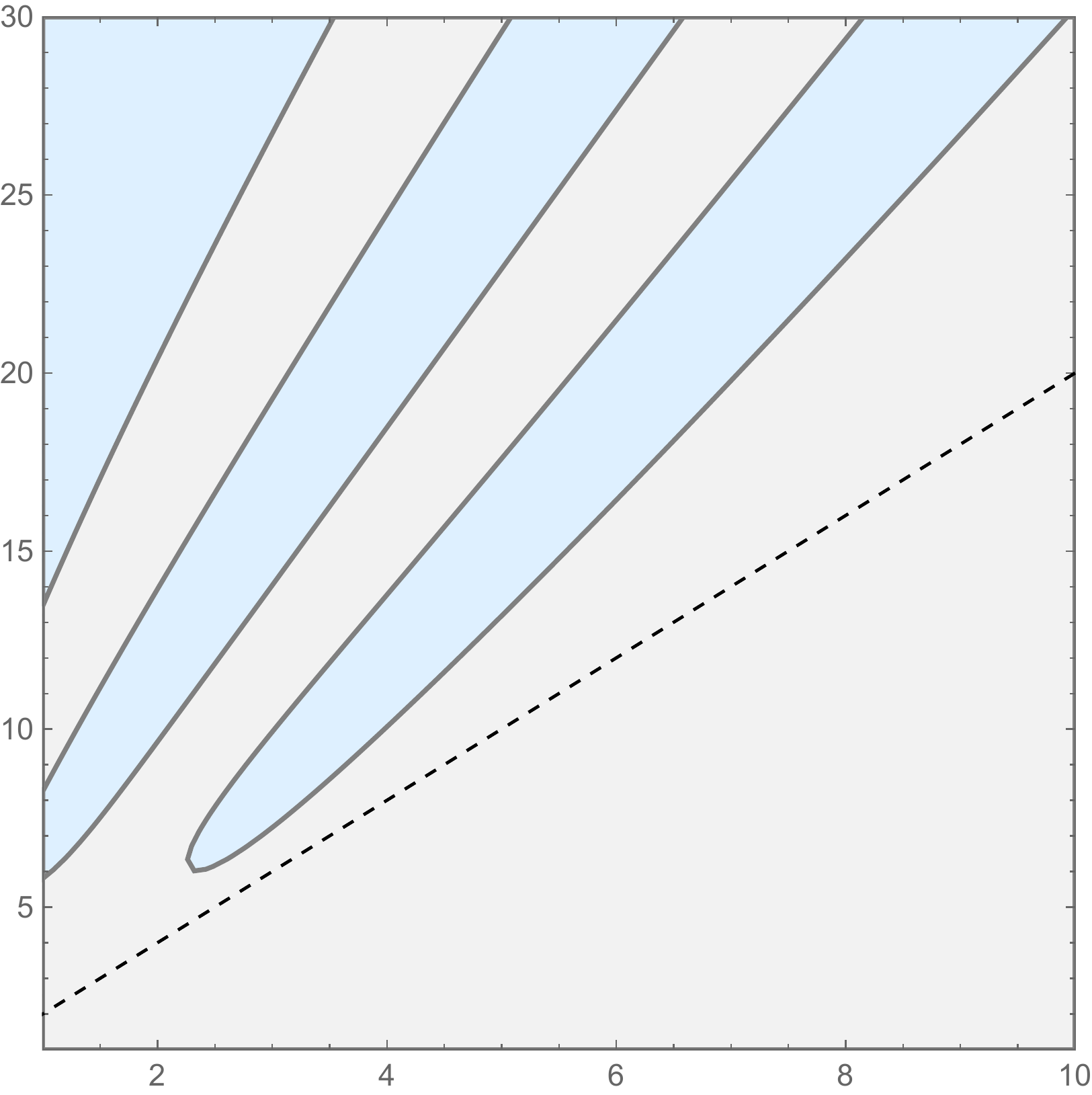}}
	\end{subfigure}

\end{array}$

\caption{Similar regions as in Figure \ref{Fig:newbound4d}, but for higher derivatives of $\tilde{F}$. Gray areas are the ones with ``correct" signed derivative in each case (negative for odd derivatives and positive for the even ones). The blue regions are those with ``wrong" signed derivatives.}

\label{Fig:higherderivatives}
\end{figure}

Until now, we have focused in the first derivative linear analysis, but as we have said before, the complete RP analysis involves all derivatives. We can repeat the linear analysis for higher derivatives. The plots for derivatives from order two to five can be seen in Figure \ref{Fig:higherderivatives}. From this linear analysis, we observe that no restriction comes from even derivatives. This is because it is expected a CFT theory to have exchange fields of high conformal dimensions in its spectrum that compensate for the mismatch in the behavior as $x\rightarrow{0}$ at both side of the CS equation (See for example section 3.1 in \cite{lightconebootstrap}). As all the upper regions for the even derivatives are gray, this means that there will always be an exchange field with a correct signed even derivative. So no restrictions will come from even derivatives.\footnote{Higher order derivatives involves very intensive time consuming calculations, but we don't expect any change of behavior for derivatives higher that the ones showed at Figure \ref{Fig:higherderivatives}.} On the other hand, for odd derivatives, we can state that \textit{there must be at least one exchange field below the upper blue region in the gray areas of each odd derivatives}. The positivity conditions for each derivative are independent. For example, there could be only one exchange field at the intersection of all gray areas to assure the correct sign of all derivatives, but that is not necessary.

We didn't find yet a wise way to make the analysis for arbitrary higher derivatives. We think that stronger constraints will come from the non linear local analysis and the global analysis using the asymptotic expansion of the next section.

It is interesting to compare this RP analysis with the bound for the minimum needed conformal dimension of the exchange fields founded originally in \cite{RychkovBound}, which was improved in \cite{RychkovBound2}:

$$\Delta \leq 2+0.7(\Delta_0-1)^{1 / 2}+2.1(\Delta_0-1)+0.43(\Delta_0-1)^{3 / 2}$$

This limiting curve, valid for $\Delta_0=$ near $1$ (the dimension of the free scalar field in $d=4$) is always above the dotted line of Figure \ref{Fig:newbound4d} and below our limiting curve. This inequality says that it should exist a non vanishing OPE coefficient corresponding to a conformal dimension of an spinless exchange field satisfying that bound.

As we have anticipated, the restriction obtained by the sum rule is stronger that the one obtained here.

\subsection{Asymptotic expansion}\label{asymtoticexpansion}

Let us now discuss the positivity using the second approach, based in demanding that the expansion in powers of $e^{-\lambda\rho}$ will have positive coefficients. We meet with some issues here that we will face in a future work. The main difficult is in the fact that, besides terms of the form $e^{-\lambda\rho}$, we also have terms like $\rho e^{-\lambda\rho}$. We will refer to these terms as ``logarithmic terms" because they came from the logarithmic terms present in the asymptotic expansion of the hypergeometric functions of the conformal blocks (See the $\log$ terms in equations (\ref{sigma3}) and (\ref{sigmap3}) of the appendix).

In the conformal blocks (for any even spin) it appears hypergeometric functions of the form:

\begin{equation}\label{HYP_CB}
	{ }_{3}F_{2}\left[\begin{array}{l}
		\frac{\Delta}{2}, \frac{\Delta}{2} + r, \frac{\Delta}{2} - \varepsilon \\
		\frac{\Delta}{2}+r+\frac{1}{2}, \Delta - \varepsilon
	\end{array} ; Z\right],
\end{equation}
with $r$ an integer number from $0$ to $n$ ($\ell =2n$ for even spin $\ell$, $n \in \mathbb{N}$) and $Z=\frac{x^2}{-4(1-x)}$, being $x$ the cross-ratio, that after the exchange $x \rightarrow 1-x$, becomes $Z_R=\frac{(1-x)^2}{-4x}$ for the rhs.

What we need is an expansion for large value of $\rho$, which means $Z_R$ large. So, we need an expansion valid for $Z_R\in\;(\infty,-1)$. In appendix \ref{expdodd}, we discuss how we get this expansion. 

 We want to consider how the term $\rho e^{-\lambda\rho}$ spoils the complete monotonicity and how it could be canceled once the infinite number of exchanges fields are included.\footnote{Notice that these type of terms comes from the logarithmic terms shown in the appendix. Schematically, the first part of the $k$th-term in (\ref{sigma3}) and (\ref{sigmap3}) becomes $Z^{-k}_R \log(-Z_R) \sim \sum_{\lambda, \beta} \rho e^{-\lambda\rho} + b e^{-\beta\rho}$.}

Let us analyze the function $f(\rho)=\rho e^{-\lambda\rho}$ in the interval $(\rho_0,\infty)$ , with $\rho_0$ some positive value. One can see that this is a non CM function, since the derivative of order $n$ has the wrong sign for $\rho < \frac{n}{\lambda}$. Despite this, one can expect that the sum of several terms like this one, together with terms like $e^{-\lambda\rho}$, with different coefficients and different values of $\lambda$, will be combined to give a CM function.

One can prove that a sum of a finite number of such terms can not give a CM function. However, it is known that there will be an infinite number of primary both side of the CS equation. This was pointed out several times in the literature. See for instance the comments on section 10.3 in \cite{TasiDuffin} or section 3.1 in \cite{lightconebootstrap}. The argument is based on the mismatching in the behavior as $x\rightarrow{0}$ in both side of the CS equality in each conformal block. We encounter here the need of the infinite number of primaries, motivated by the impossibility of having a CM function in the rhs considering only a finite numbers of conformal blocks.

It will be interesting to know then, under what conditions of the CFT data the infinite sum in the rhs of the CS equation becomes a CM function, i.e. an expansion of only terms like $e^{-\lambda\rho}$ with positive coefficients. We leave this for a future work.

\subsection{Case of non equal external spinless fields}

Now, one can consider the more interesting case in which the 4-point function involves two different spinless field of dimensions $\Delta_1$ and $\Delta_2$. The reason why this is a more interesting case comes from the fact that different structure constants appear on both sides of the equality:

\begin{equation}
\frac{1}{x^{\Delta_1+\Delta_2}}\sum_p (-1)^{\ell_p} C_{12}^2 {F^{21}_{21}}^{(p)}(x)= \frac{1}{(1-x)^{2\Delta_2}}\sum_p C_{11}^p C_{22}^p {F^{11}_{22}}^{(p)}(1-x)
\end{equation}
where $\ell_p$ is the spin of the exchange field\footnote{One can be worried about the positivity of the coefficients on the lhs, but the minus sign for odd spin exchange fields coming from $(-1)^{\ell_p}$ is compensated with an equal one coming form the conformal blocks ${F^{21}_{21}}^{(p)}(x)$}, and is to be noticed that in the rhs only exchange fields with even spin appear.

One can see that in the right hand side the OPE coefficients appears in the form $ C_{11}^p C_{22}^p$, which is not a priori positive. Therefore, in the subsequent analysis, we need to take into account the unknown sign of this combination in addition to the conformal dimension and spin which labels the conformal blocks. This introduce more freedom, making the constrains weaker.

When one considers CFT in $d$ greater than 2, since the ${F^{ba}_{cd}}^{(p)}$ depend only on the difference $\Delta_a-\Delta_b$ and $\Delta_{c}-\Delta_{d}$, in the rhs we have a simple (with known closed form conformal blocks) version of the CS equation:

\begin{equation}
	\frac{1}{x^{\Delta_1+\Delta_2}}\sum C_{12}^2{F^{21}_{21}}^{(p)}(x)= \frac{1}{(1-x)^{2\Delta_2}}\sum C_{11}^p C_{22}^p F^{(p)}(1-x),
\end{equation}

where by ``simple" we mean that now, $F^{(p)}$ in the rhs is known in closed form as it is the same conformal block seen before which appears in the case in which all external fields are identical.

Now, we can repeat the previous procedure leading to the CM character of :

\begin{equation}
	\frac{x^{\Delta_1+\Delta_2}}{(1-x)^{2\Delta_2}-x^{\Delta_1+\Delta_2}}\sum_p C_{11}^p C_{22}^p F^{(p)}(1-x)
\end{equation}

when $x$ is such that $(1-x)^{2\Delta_2}-x^{\Delta_1+\Delta_2}>0$. These conditions hold for $x$ less than certain $x_0$, determined by the condition $(1-x)^{2\Delta_2}=x^{\Delta_1+\Delta_2}$

In contrast with the approach based on the sum rule, here one can derive inequalities for a quantity involving only $C_{11}^p C_{22}^p$ and not $C_{12}^p$, and as only the rhs is involved in the constrains, then only the functions for even spins $F^{(p)}$ (which are available in closed form) are used. Also, there is no need to calculate the conformal blocks present in the lhs that are only known from recurrence relations and more computational power is required to calculate them and its derivatives numerically.

It is important to remark that in this case, when the external conformal dimensions are not equal, the $\eta-$inequalities are not implied by the $\rho-$types. So, both types of inequalities should be considered. We left for a future work the study of this case.

\section{Final remarks}

 We have introduced a reflection positivity approach that can be thought of as a complementary tool to the main one based on the sum rule \cite{RychkovBound}. The main point of this paper was to show that RP itself is a strong condition which contains an important part of the restrictions imposed by the full CS equality (\ref{CS}). The key observation was that, in contrast with the lhs of eq (\ref{CS}), the rhs does not fulfill automatically RP for an arbitrary choice of OPE coefficients. 
 
We have started our discussion with the case in which all the external fields (scalars fields) have the same conformal dimensions. When there is a pair of non-equal conformal dimensions, this RP based approach becomes more useful since it leads to constrains involving only the OPE coefficients and the conformal blocks (which are known in closed form) of the rhs. In other word, in this case the rhs becomes considerable simpler (with conformal blocks in closed form and absence of odd spin of exchange field) and this approach allows to disentangling the lhs from this.

One important aspect of this approach was the use of simple inequalities arising by the applications of RP to one-dimensional arrange of points, used by us in \cite{Blanco:2019gmt}. Instead of using the full RP inequalities, we had to deal only with positive definite functions of a single variable and, in particular, with complete monotonic functions of a variable related with the real cross-ratio. Two different arrange of points lead to $\rho=-\log(x)$ and $\eta=\frac{1}{\sqrt{x}}-1$ as suitable parameters.

As we have mentioned in section \ref{generalproc}, the complete monotonic character of a function can be expressed in three different ways. We have focused mostly on the linear inequalities, which say that all even (odd) derivatives of rhs, as function of $\rho=-\log(x)$ (an also of the variable $\eta$), should be positive (negative). We expect that a more exhaustive use of the RP positivity on the rhs, i.e., the use of the asymptotic expansion in $e^{-\rho}$ and/or the use of the non-linear inequalities, will lead to an improved version of the constraints described here. A good illustration of the power of the non-linear analysis vs the linear one, can be seen in this simple example: 

$$f(x)=(50+x)e^{-x}$$

This is a non complete monotonic function. However, if one analyze only the signs of its derivative, would not realize this until computing the 51-th derivative!. In contrast, the computation of determinant (\ref{det1}) of order $N=2$ gives: $-e^{-2x}$ which immediately shows that $f$ is a non cm function. These type of functions appears in the asymptotic expansion analyzed in section \ref{asymtoticexpansion}.

As a by product, we have seen how RP applied on the lhs (where RP is automatically fulfilled) can be useful for uncovering positivity conditions of the conformal blocks which could not be evident. We see how RP applied to different configurations of points lead to positivity conditions for coefficients appearing in the conformal block multiplied for different factors, depending on the arrangements under consideration.

The inequalities of $\eta$-type are the less obvious and go beyond the positivity of the coefficients in the $x$ expansion that was first mentioned in \cite{Hartman:2015}. It is important to recall that $\eta$ variable has the meaning of a distance, as is displayed in Figure \ref{Fig:sequences}, and the corresponding inequalities are also valid for a 4-point of a general QFT.

\section*{Acknowledgments}

We thank the members of the hep-th group of Physics department of University of Buenos Aires and specially to G. Giribet for fruitful discussions. This work was supported by National Scientific and Technical
Research Council (CONICET) and University of Buenos Aires.

%\section*{Appendix}
% \addcontentsline{toc}{section}{Appendix}

\appendix

\section{Conformal Block asymptotic expansion}\label{expdodd}
\addtocontents{toc}{\protect\setcounter{tocdepth}{1}}
In \cite{Diagonal}, it has been found the closed form of the conformal blocks in any dimension ($d$), for the case of four external scalars with $\Delta_{12}=0$, $\Delta_{34} \neq 0$. In general, if we have two different scalar fields in the lhs of the CS equation $\left< \phi_1 \phi_2 \phi_2 \phi_1 \right>$, then the rhs is of the form $\left< \phi_1 \phi_1 \phi_2 \phi_2 \right>$ and we get the case $\Delta_{12}=\Delta_{34}=0$, with only even spins present in the OPE. As seen in \cite{Diagonal}, in these conformal blocks it appears hypergeometric functions of the form:

\begin{equation}\label{HYP_CB}
	{ }_{3}F_{2}\left[\begin{array}{l}
		\frac{\Delta}{2}, \frac{\Delta}{2} + r, \frac{\Delta}{2} - \varepsilon \\
		\frac{\Delta}{2}+r+\frac{1}{2}, \Delta - \varepsilon
	\end{array} ; Z\right],
\end{equation}
with $r$ an integer number from $0$ to $n$ ($\ell =2n$ for even spin $\ell$, $n \in \mathbb{N}$), $\epsilon=\frac{d-2}{2}$ and $Z=\frac{x^2}{-4(1-x)}$, being $x$ the cross-ratio, that after the exchange $x \rightarrow 1-x$, becomes $Z_R=\frac{(1-x)^2}{-4x}$ for the rhs.

It can be seen that the relevant information in the CS equation is in the range $\rho \in [\rho_c, +\infty )$, for some arbitrary finite value $\rho_c$, and then by the completely monotonic and analytic properties of the CS eq. the rest will be fulfilled.\footnote{In fact, it can be seen from \cite{CMintegers}, that from the properties of completely monotonic functions, it is sufficient to impose the equality of lhs and rhs of the CS equation only at $\rho_c$ integers from an arbitrary number $N$ to $+\infty$, for the equality to be satisfied at all $\rho$ real numbers (or at all $x$ reals in $(0,1)$)}
So, it is very important to have the expansion of the hypergeometric function (\ref{HYP_CB}) when $|Z|>1$, that is for $Z_R$ in the RHS when $\rho_c \in \left( -\log (3-2\sqrt{2}),\ +\infty \right)$ (or $x\in \left( 0,\ 3-2\sqrt{2}\right)$).

For general $Z$ values the analytically continued hypergeometric function ${}_p F_q$ have an integral representation, when $Z \neq 0$ and $a_{k} \neq 0,-1,-2, \ldots$, for $k=1,2, \ldots, p$:

\begin{equation}
	\begin{array}{l}
		\displaystyle\left(\prod_{k=1}^{p} \Gamma\left(a_{k}\right) / \prod_{k=1}^{q} \Gamma\left(b_{k}\right)\right) {{}_{p}F_{q}}\left(\begin{array}{l}
			a_{1}, \ldots, a_{p} \\
			b_{1}, \ldots, b_{q}
		\end{array}; Z\right) \\
		\\
		\displaystyle\quad=\frac{1}{2 \pi i} \int_{c-i\infty}^{c+i\infty} \left(\prod_{k=1}^{p} \Gamma\left(a_{k}+s\right) / \prod_{k=1}^{q} \Gamma\left(b_{k}+s\right)\right) \Gamma(-s)(-Z)^{s} \mathrm{~d} s,
	\end{array}
\end{equation}
where $\Gamma(\cdot)$ is the Gamma function, and the contour of integration, for example for the $|Z|>1$ case, needs to be closed at the infinity of the left half plane, separating the poles of $\Gamma(-s)$ and those generated by the factor in parenthesis with the other $\Gamma$ functions.

When $a_{k} = 0,-1,-2, \ldots$ for some $k$ in $1,2, \ldots, p$, regardless of the value of $Z$, the hypergeometric function is a polynomial of finite order, defined by the usual hypergeometric series that get cut at $N=min(|a_k|$ \ / $a_k \in -\mathbb{N}_0)$, i.e., the minimal value between the modules of the $a_k$'s that are negative integers,
\begin{equation}\label{negativeintegers}
	{{}_{p}F_{q}}\left(\begin{array}{l}
		a_{1}, \ldots, a_{p} \\
		b_{1}, \ldots, b_{q}
	\end{array}; Z\right)= \sum_{n=0}^{N} \frac{\prod_{k=1}^{p} \left(a_{k}\right)_n}{\prod_{k=1}^{q} \left(b_{k}\right)_n} \frac{Z^n}{n!}
\end{equation}

%\subsection[Odd ${d}$ expansion]{Odd $\boldsymbol{d}$ expansion}

Next, we will limit only to the \textbf{odd} $\boldsymbol{d}$ case.\footnote{For even $d$, the expansion is different, we are not showing it here. For example it also has terms of order $\log^2 (-Z)$.}

In the conformal block, we need to expand the hypergeometric ${}_3 F_2$ (\ref{HYP_CB}). If we define $T$ as the expansion of the hypergeometric: 
\begin{equation}
	\begin{array}{l}
		\displaystyle T\equiv \frac{\Gamma\left(\frac{\Delta}{2}\right) \Gamma\left(\frac{\Delta}{2} + r\right) \Gamma\left(\frac{\Delta}{2} - \varepsilon\right)}{\Gamma\left(\frac{\Delta}{2}+r+\frac{1}{2}\right) \Gamma\left(\Delta - \varepsilon\right)}{ }_{3}F_{2}\left[\begin{array}{l}
			\frac{\Delta}{2}, \frac{\Delta}{2} + r, \frac{\Delta}{2} - \varepsilon \\
			\frac{\Delta}{2}+r+\frac{1}{2}, \Delta - \varepsilon
		\end{array} ; Z\right] \\
		\\
		\displaystyle\quad=\frac{1}{2 \pi i} \int_{c-i\infty}^{c+i\infty} \frac{\Gamma\left(\frac{\Delta}{2} +s\right) \Gamma\left(\frac{\Delta}{2} + r +s\right) \Gamma\left(\frac{\Delta}{2} - \varepsilon +s\right) }{\Gamma\left(\frac{\Delta}{2}+r+\frac{1}{2} +s\right) \Gamma\left(\Delta - \varepsilon +s\right)} \Gamma(-s)(-Z)^{s} \mathrm{~d} s,
	\end{array}
\end{equation}
and take into account unitary bound, that $r$ is an integer (from $0$ to $n$) and $\epsilon=\frac{d-2}{2}$ a half integer for odd $d$, the expansion $T$ becomes a sum of terms coming from the different poles of the first factor of the integral representation, that for odd $d$, differs if $\Delta$ is an odd integer or not. We present these cases next.

\subsection[$\Delta \neq 2\mathbb{N}-1$]{$\boldsymbol{\Delta \neq 2\pmb{\mathbb{N}}-1}$}

$T=\Sigma_1 + \Sigma_2 + \Sigma_3$

\begin{equation}
	\Sigma_1=\frac{(-Z)^{-\frac{\Delta }{2}} \Gamma \left(\frac{\Delta }{2}\right) \Gamma (-\varepsilon ) \Gamma (r) }{\Gamma \left(r+\frac{1}{2}\right) \Gamma \left(\frac{\Delta }{2}-\varepsilon \right)}\sum _{k=0}^{r-1} \frac{ \left(\frac{\Delta }{2}\right)_k \left(\frac{1}{2}-r\right)_k \left(-\frac{\Delta }{2}+\varepsilon +1\right)_k}{ (\varepsilon +1)_k (-(r-1))_k}\frac{Z^{-k}}{k!}
\end{equation}
(with $\Sigma_1=0$ if $r=0$).

\begin{equation}
	\Sigma_2=\frac{(-Z)^{\varepsilon -\frac{\Delta }{2}} \Gamma (\varepsilon ) \Gamma \left(\frac{\Delta }{2}-\varepsilon \right) \Gamma (r+\varepsilon ) }{\Gamma \left(\frac{\Delta }{2}\right) \Gamma \left(r+\varepsilon +\frac{1}{2}\right)} \, { }_{3}F_{2}\left[\begin{array}{l}
		\frac{\Delta }{2}-\varepsilon, -r-\varepsilon +\frac{1}{2}, 1-\frac{\Delta }{2} \\
		1-\varepsilon, -r-\varepsilon +1
	\end{array} ; \frac{1}{Z}\right]
\end{equation}

Note that the ${}_3 F_2$ hypergeometric in $\Sigma_2$ has the usual expansion as $|1/Z|<1$, and in fact, it is a polynomial, as the sum gets cut because at least one of its upper arguments is always a negative integer or zero.

\begin{equation}\label{sigma3}
	\begin{aligned}
		\Sigma_3=&\frac{(-1)^r (-Z)^{-\frac{\Delta }{2}-r} \Gamma \left(r+\frac{\Delta }{2}\right) \Gamma (-r-\varepsilon ) }{\Gamma \left(\frac{1}{2}\right) \Gamma \left(-r+\frac{\Delta }{2}-\varepsilon \right) r!}\\
		&\sum _{k=0}^{+\infty} \frac{ \left(\frac{1}{2}\right)_k \left(r+\frac{\Delta }{2}\right)_k \left(r+\varepsilon -\frac{\Delta }{2}+1\right)_k }{ (r+1)_k (r+\varepsilon +1)_k} \left[\log (-Z)+h_{2,k}\right] \frac{Z^{-k}}{k!},
	\end{aligned}
\end{equation}
where
\begin{equation}
	\begin{aligned}
		h_{2,k}=-&\psi (-k-r+\frac{\Delta }{2}-\varepsilon )-\psi (k+r+\frac{\Delta }{2})+\psi (-k-r-\varepsilon )\\
		+&\psi (k+r+1)+\psi (k+1)-\psi (\frac{1}{2}-k),
	\end{aligned}
\end{equation}

and $\psi$ is the Digamma function.

\subsection[$\Delta = 2\mathbb{N}-1$$\iff$$\frac{\Delta}{2} - \varepsilon \in \mathbb{Z}$]{$\boldsymbol{\Delta = 2\pmb{\mathbb{N}}-1$ $\iff$ $\frac{\Delta}{2} - \varepsilon \in \pmb{\mathbb{Z}}}$}
In this case we have two possibilities:

$\bullet$ { $\frac{\Delta}{2} - \varepsilon \geq r$ (if $r \neq 0$) and $\frac{\Delta}{2} - \varepsilon > 0$ (if $r = 0$)}

$T=\Sigma_1+\Sigma_2+\Sigma'_3+\Sigma_4$

\begin{equation}\label{sigmap3}
	\begin{aligned}
		\Sigma'_3=&\frac{(-1)^r (-Z)^{-\frac{\Delta }{2}-r} \Gamma \left(r+\frac{\Delta }{2}\right) \Gamma (-r-\varepsilon ) }{\Gamma \left(\frac{1}{2}\right) \Gamma \left(-r+\frac{\Delta }{2}-\varepsilon \right) r!}\\
		&\sum _{k=0}^{\frac{\Delta }{2}-\varepsilon -r-1} \frac{ \left(\frac{1}{2}\right)_k \left(r+\frac{\Delta }{2}\right)_k \left(r+\varepsilon -\frac{\Delta }{2}+1\right)_k }{ (r+1)_k (r+\varepsilon +1)_k} \left[\log (-Z)+h_{2,k}\right] \frac{Z^{-k}}{k!},
	\end{aligned}
\end{equation}

(with $\Sigma'_3=0$ if $\frac{\Delta }{2}-\varepsilon -r=0$).

Note that $\Sigma'_3$ is the same as $\Sigma_3$, but the sum upper limit now is finite.

\begin{equation}
	\begin{aligned}
		\Sigma_4=&\frac{(-1)^r (-Z)^{\varepsilon -\Delta } \Gamma \left(-\frac{\Delta }{2}\right) \Gamma (\Delta -\varepsilon ) }{\left(\frac{\Delta }{2}-\varepsilon \right)! \left(\frac{\Delta }{2}-\varepsilon -r\right)! \Gamma \left(r+\varepsilon -\frac{\Delta }{2}+\frac{1}{2}\right)}\\
		&\quad \quad { }_{4}F_{3}\left[\begin{array}{l}
			1,1,\Delta -\varepsilon ,-r+\frac{\Delta }{2}-\varepsilon +\frac{1}{2} \\
			\frac{\Delta }{2}-\varepsilon +1,-r+\frac{\Delta }{2}-\varepsilon +1,\frac{\Delta }{2}+1
		\end{array} ; \frac{1}{Z} \right]
	\end{aligned}
\end{equation}

Note that the ${}_4 F_3$ hypergeometric has the usual expansion as $|1/Z|<1$.

$\bullet$ { $\frac{\Delta}{2} - \varepsilon \leq 0$}

This case is only possible for r=0, and $T$ is just the regular hypergeometric expansion (\ref{negativeintegers}), resulting in a polynomial of order $N=-(\frac{\Delta}{2} - \varepsilon)$, i.e., an expansion of positive powers of $Z$.\footnote{For example, for $d=3$ the hypergeometric becomes just 1, as for $d=3$, $\Delta=1$ is the only odd integer possible in this case.}

\bibliographystyle{toine}
\bibliography{Refs}{}

\end{document}